\documentclass{aa}
\usepackage[varg]{txfonts}
 
\usepackage{lipsum}
\usepackage{graphicx}
\usepackage{txfonts}
\usepackage{xcolor}
\usepackage{orcidlink}

\usepackage{siunitx}
\DeclareSIUnit\parsec{pc}
\DeclareSIUnit\gauss{G}
\DeclareSIUnit\erg{erg}
\DeclareSIUnit\year{yr}
\DeclareSIUnit\arcmin{arcmin}
\DeclareSIUnit\arcsec{arcsec}
\DeclareSIUnit\counts{cts}
\DeclareSIUnit\jansky{Jy}
\DeclareSIUnit\steradian{sr}

\newcommand{\hi}{H\,{\sc i}\xspace}
\newcommand{\cm}{cm$^{-2}$\xspace}
\newcommand{\HI}{H\,{\sc i}\xspace}
\newcommand{\kms}{km~s$^{-1}$\xspace}
\newcommand{\mJybeam}{mJy beam$^{-1}$\xspace}

\newcommand{\msun}{\ensuremath{M_{\odot}}\xspace}

\newcommand{\degree}{\ensuremath{^\circ\xspace}}

\newcommand{\rzero}{{\tt r00\_t00}\xspace}

\newcommand{\erwin}[1]{{\color{black} #1}}

\usepackage{ctable}

\hypersetup{
    colorlinks=true,
    linkcolor=black,
    citecolor=blue,
    urlcolor=blue
}

\begin{document}

    \title{\HI absorption in MHONGOOSE}
    \subtitle{Spin temperatures and cold neutral medium in nearby disk galaxies}

    \titlerunning{HI absorption in MHONGOOSE}

        \author{W.J.G. de Blok\inst{\ref{inst1}, \ref{inst2}, \ref{inst3}}\orcidlink{0000-0001-8957-4518}
        \and
            F.M. Maccagni\inst{\ref{inst4}, \ref{inst5}}
        \and
            L. Chemin\inst{\ref{inst6}}
        \and
            K. Haubner\inst{\ref{inst7}, \ref{inst8}}
            \orcidlink{0009-0007-7808-4653}
        \and
            R. Morganti\inst{\ref{inst1},\ref{inst3}}
        \and
            T.A. Oosterloo\inst{\ref{inst1},\ref{inst3}}
        \and
            D.~Kleiner\inst{\ref{inst1}}
        \and
            S.~Veronese\inst{\ref{inst9}}
        \and 
            S.~Kurapati\inst{\ref{inst10}}
        \and
            J.~Healy\inst{\ref{inst11},\ref{inst2}}
        }
    
    \institute{Netherlands Institute for Radio Astronomy (ASTRON), Oude Hoogeveensedijk 4, 7991 PD Dwingeloo, The Netherlands\label{inst1}
        \and
         Department of Astronomy, University of Cape Town, Private Bag X3, Rondebosch 7701, South Africa\label{inst2}
        \and
         Kapteyn Astronomical Institute, University of Groningen, PO Box 800, 9700 AV Groningen, The Netherlands\label{inst3}
        \and 
         INAF -- Osservatorio Astronomico di Cagliari, Via della Scienza 5, I-09047, Selargius, (CA), Italy\label{inst4}
         \and
         Wits Centre for Astrophysics, School of Physics, University of the Witwatersrand, 1 Jan Smuts Avenue, 2000, Johannesburg, South Africa\label{inst5}
         \and
         Universit\'e de Strasbourg, CNRS, Observatoire astronomique de Strasbourg, UMR 7550, 67000 Strasbourg, France\label{inst6}
         \and
         INAF -- Arcetri Astrophysical Observatory, Largo Enrico Fermi 5, 50125, Firenze, Italy\label{inst7}
         \and 
         Dipartimento di Fisica e Astronomia, Universit\`a degli Studi di Firenze, Via G. Sansone 1, 50019, Sesto Fiorentino, Firenze, Italy\label{inst8}
         \and
         Max Planck Institute for Radio Astronomy, Auf dem Hügel 69, 53121 Bonn, Germany\label{inst9}
         \and
         National Centre for Radio Astrophysics, Tata Institute of Fundamental Research, Pune University Campus, Post Bag 3, Ganeshkhind Pune 411007, India\label{inst10}
         \and
         Jodrell Bank Centre for Astrophysics, School of Physics and Astronomy, University of Manchester, Oxford Road, Manchester M13 9PL, UK\label{inst11}
        }

    \date{Received [Month day, year]; accepted [Month day, year]}


\abstract{
Combined \hi emission--absorption studies constrain the spin temperature and phase structure of the neutral atomic hydrogen interstellar medium (ISM), but have largely been limited to the Milky Way and the Local Group. We extend this technique to galaxies at distances of 7--22\,Mpc using deep data from the MeerKAT \HI Observations of Nearby Galactic Objects -- Observing Southern Emitters (MHONGOOSE) survey, and quantify the detection fraction and Cold Neutral Medium (CNM) properties at these distances.
We search for \hi absorption toward 56 background continuum sources in 21 out of the 30 MHONGOOSE galaxies (with nine galaxies lacking suitable background sources), and detect absorption associated with the galaxies' \HI disks in three cases: one sight line in NGC~289 and two in NGC~7424. This corresponds to detection rates of 3/56 (5 percent) for the full sample and 3/31 (10 percent) for a 
clean sub-sample of sight lines, considering only unresolved background sources behind 14 low-inclination galaxies.
Detections occur only where both the continuum flux and the foreground \HI column density are high, with optical-depth sensitivity as the primary limiting factor.
For the detected sight lines, we model the absorption and emission spectra to derive spin temperatures and CNM fractions using the standard combined emission--absorption method. The CNM spin temperatures and line widths are comparable to Local Group measurements, but the inferred CNM fractions are systematically lower.
We argue that this difference is primarily a resolution effect: at the distances of our galaxies, the emission spectra average over several hundred parsecs, diluting structured CNM relative to the smoother Warm Neutral Medium (WNM). This demonstrates that emission--absorption analyses can be extended beyond the Local Group, provided that care is taken in constructing representative emission spectra.
}


\maketitle

\section{Introduction}

Neutral atomic hydrogen (\hi) is the dominant baryonic component of the interstellar medium (ISM) in most 
disk galaxies. It provides the reservoir from which molecular gas and stars form. 
Observations of the \hi 21-cm radio line play an important role in studies of galaxy structure, kinematics, and evolution.

The atomic ISM can be described in terms of two phases in approximate pressure equilibrium 
\citep{field1969,wolfire1995,wolfire2003, mccluregriffiths2023}.
The cold neutral medium (CNM) has characteristic temperatures of ${\sim}10$\,K to ${\sim}100$\,K, while the warm neutral medium (WNM) has temperatures of ${\sim}10^4$\,K. Quantifying how atomic gas is distributed between these phases, and how this distribution varies with galaxy properties or environment, remains a key goal in ISM studies.
The CNM has a lower volume-filling factor and represents the atomic phase most directly connected to molecular clouds and star formation. The WNM is more diffuse, dominates the volume, and  traces the larger-scale distribution of the ISM. 

The spin temperature $T_s$ is the excitation temperature that describes the relative population of the two hyperfine levels of the ground state of the neutral hydrogen atom and provides a direct constraint on the physical state of the \hi. It is set by a competition between collisional coupling, Ly$\alpha$ scattering, and the ambient radiation field. In the CNM, the high densities ensure that collisions dominate, so that the spin temperature approximates the kinetic temperature. In the WNM, the collision process is inefficient due to the lower densities, and $T_s$ is instead determined by a combination of Ly$\alpha$ coupling and the background radiation field, resulting in values of a few times $10^3$\,K  that are generally lower than the kinetic temperature \citep{liszt2001}.  

Most studies of galaxies in \hi are based on 21-cm emission which is produced by both the WNM and CNM. 
Emission by itself provides only limited information on the phase structure of the ISM. For example, cold gas can become optically thick, affecting the proportionality between measured emission and column density, narrow CNM emission components can blend with more dominant and broader WNM components along the line of sight, and emission measurements can be affected by beam dilution when the angular scale of \hi structures is smaller than the telescope beam. Since the CNM is often organized into compact structures, this can bias emission-based estimates of its properties.

\hi absorption offers a complementary constraint. 
Due to the higher absorption efficiency of cold gas, set by the hyperfine level populations of \HI and the inverse dependence of the optical depth on spin temperature, absorption against a background radio continuum source primarily traces the cold component of the foreground gas, i.e. the CNM.
When absorption and emission measurements are combined, it becomes in principle possible to constrain the CNM contribution to the emission spectrum. One complication is that absorption measurements probe a narrow, pencil-beam line of sight, while emission measurements are averaged over the telescope resolution and usually taken at positions adjacent to the absorber.
The underlying assumption that both types of measurements trace the same structures may not always be valid and this becomes increasingly critical at larger distances.

The most detailed combined emission--absorption measurements can be carried out in the MW, owing to both the large number of bright background continuum sources and the high physical resolution of the observations. \HI surveys (e.g., \citealt{dickey1990,heiles2003a,heiles2003b,dickey2003,murray2015,murray2018,murray2021,mccluregriffiths2023}) have shown that emission and absorption spectra often differ substantially, and that these differences can be explained by the coexistence of cold and warm gas along the line of sight. These studies show that the CNM constitutes a significant fraction of the Galactic \hi, with a broad distribution of spin temperatures and a preference for values between ${\sim}40$\,K and ${\sim}70$\,K. \citet{kanekar2011} used absorption measurements in the MW to identify a rapid change in spin temperature at a column density of ${\sim}2 \times 10^{20}$ \cm, interpreted as the threshold for CNM formation.

The Magellanic Clouds (MCs) are the closest environment outside the MW where the phases of the ISM can be studied under different physical conditions, including lower metallicity and stronger radiation fields. The MCs generally have a lower CNM fraction compared to the MW (e.g., \citealt{stanimirovic1999,dickey2000,jameson2019,dempsey2022}). 
Absorption detections in M31 and M33 have been used to measure the properties of cold atomic gas in other Local Group (LG) galaxies (e.g., \citealt{dickey1988,dickey1993,braun1992}). \citet{pingel2024} reported absorption detections in the LG dwarf galaxy NGC~6822 and derived spin temperatures and CNM fractions.

Outside the LG (at distances beyond ${\sim}1$\,Mpc), background continuum sources such as active galactic nuclei (AGN) or quasars are not only used as probes but can themselves be the targets of absorption studies. In many cases, the absorption arises in the AGN itself rather than in an intervening galaxy. This associated absorption probes the circumnuclear environment. Central regions of AGN typically do not show cold gas in \hi emission. However, about 30 percent of them show cold-gas absorption, allowing the cold components to be traced \citep{maccagni2017}. In many cases, this absorption traces inflows and outflows interpreted as jet--ISM interaction \citep{morganti2018}. 

In the case of intervening absorption, the absorbing gas is unrelated to the emitting source and instead probes neutral gas in galaxy disks and halos along the line of sight between the observer and the background emitter. 
Many studies of intervening absorption have been guided by searches for Damped Lyman-$\alpha$ (DLA) and Mg\,{\sc ii} absorbers, which in many cases have defined the samples for 21-cm absorption follow-up (e.g., \citealt{lane2000,kanekar2003,kanekar2009}). It is not clear to what extent this pre-selection introduces biases in, for example, metallicity or dust content.

Wide-field \hi absorption surveys conducted with Square Kilometre Array (SKA) pathfinder and precursor telescopes should help mitigate these biases by avoiding selection based on optical absorption. Two of these surveys are the MeerKAT Absorption Line Survey (MALS; \citealt{gupta2016}) and the First Large Absorption Survey in \hi  (FLASH; \citealt{allison2022}). MALS is a deep, unbiased search for \hi and OH absorption over the redshift range $0 < z < 2$, targeting both intervening absorbers and associated absorption associated with radio-loud AGN. FLASH aims to detect several hundred intervening and associated absorbers over $0.4 < z < 1.0$ toward bright radio continuum sources using wide-field observations.

These surveys provide a census of cold atomic gas at intermediate redshifts that are difficult to access via \hi emission. Detailed, spatially resolved studies of nearby systems with absorption can help interpret, among other things, the observed detection rates, the distribution of impact parameters, the phase balance of the ISM, and any dependence on galaxy type.

Absorption-based constraints on the CNM and WNM in the local universe outside the LG remain limited to a small number of systems and sight lines.
For example, early observations of galaxy-quasar pairs are presented by \citet{carilli1992}, with additional examples of intervening absorption in nearby galaxies reported by \citet{borthakur2014,dutta2016,gupta2018}. \citet{maina2022} detected absorption in the Klemola~31 group at $z=0.029$, associated with the \hi disk of a group member. In a study of 16 nearby gas-rich galaxies comprising 24 sight lines, \citet{reeves2015,reeves2016} detected absorption along one line of sight. Many of these studies are limited by angular resolution, complicating the comparison between absorption and emission spectra.

In this paper, we use data from the MeerKAT \hi Observations of Nearby Galactic Objects: Observing Southern Emitters (MHONGOOSE\footnote{\tt https://mhongoose.astron.nl/}; \citealt{deblok2024}) survey to study intervening absorption in nearby galaxies. MHONGOOSE provides deep, high-resolution \hi observations of disk and dwarf galaxies, probing one to two orders of magnitude deeper in column density than previous surveys. We analyse spectra along sight lines toward these galaxies to quantify the detection fraction and the properties of the detected absorption.
Applying the combined emission--absorption analysis traditionally used in LG studies, we derive spin temperatures and CNM fractions for galaxies at distances of 7--22\,Mpc, i.e., an order of magnitude more distant than those probed in previous work.

In Sect.\ 2 we briefly describe the MHONGOOSE \HI data and the additional processing required to produce the absorption data. In Sect.\ 3 we review the necessary background on combined emission--absorption studies and introduce the equations and methodology used in the remainder of the paper. Section 4 presents the selection of the continuum sample and the corresponding spectra. Section 5 discusses the absorption detections and derives their physical properties. Section 6 examines possible reasons for the non-detections, and Sect.\ 7 summarises the paper.

\section{MHONGOOSE}

MHONGOOSE \citep{deblok2024} is a deep \hi survey of 30 nearby ($D < 23\,\mathrm{Mpc}$) disk and dwarf galaxies. Its main science goals are to study the accretion processes by which galaxies obtain sufficient gas to sustain star formation, and the relation between this gas and star formation in these systems. MHONGOOSE is the deepest interferometric \hi survey to date, with each galaxy observed for 55\,h with MeerKAT, reaching a column density sensitivity of ${\sim}5 \times 10^{17}$ \cm (3$\sigma$ over 16\,\kms) at a resolution of ${\sim}1'$--$1.5'$. For a full description of the survey parameters and data reduction procedures, see \citet{deblok2024}.

To cover the full resolution range available with MeerKAT, MHONGOOSE \hi data products were created at six ``standard'' resolutions, spanning ${\sim}7''$ to ${\sim}90''$, using different combinations of robust weighting and tapering. The highest-resolution maps are created using a robust value of zero without tapering (\rzero), yielding angular resolutions of ${\sim}7''$. At this resolution, the noise in a single 1.4\,\kms channel is 0.22\,\mJybeam, corresponding to a column density sensitivity of ${\sim}6 \times 10^{19}$ \cm (3$\sigma$ over 16\,\kms).

The continuum measurement sets of the MHONGOOSE galaxies are created during the self-calibration procedure that forms part of the standard reduction pipeline (see \citealt{deblok2024} for details). They cover the frequency range 1390--1420\,MHz. Initial imaging revealed direction-dependent artefacts in some fields, necessitating additional direction-dependent calibration, for which we used the {\tt oxkat} \citep{heywood2020} package. The final continuum images were created by combining the ten direction-dependent-calibrated measurement sets for each galaxy and imaging them using a robust weighting parameter of zero and without tapering (i.e. identical to the \rzero \HI data). We used the masks created by the {\tt oxkat} package for the deconvolution. The noise level in these continuum images is $\sim 3.3\,\mu\mathrm{Jy}\,\mathrm{beam}^{-1}$.

\section{Combining absorption and emission spectra\label{sec:absem}}

As the methodology of deriving and combining absorption and emission spectra plays a key role in interpreting our data and in defining the sample of background continuum sources, we briefly review the key equations and procedures here. Further details can be found in, e.g., \citet{dickey1992,dickey2003}, \citet{heiles2003a,heiles2003b}, \citet{murray2021}, \citet{mccluregriffiths2023} and \citet{pingel2024}.

The basic radiative transfer equation is
\begin{equation}
T_B(v) = T_s (1 - e^{-\tau(v)}),\label{eq:tau}
\end{equation}
where $T_B(v)$ is the brightness temperature at velocity $v$,
$T_s$ is the spin temperature of the gas, and $\tau(v)$ is the optical
depth. For optically thin gas ($\tau \ll 1$), we can approximate this as  
\begin{equation}
T_B(v) \approx T_s \tau(v)\label{eq:smalltau}.
\end{equation}
When \HI is observed in absorption against a background continuum source with flux density $S_{\rm cont}$,  the observed flux density as a function of velocity reflects the effect of foreground gas on the background continuum, and can be written as
\begin{equation}
S(v) = S_{\rm cont} [(1 - c_f) + c_f e^{-\tau(v)}].
\end{equation}
Here $c_f$ is the covering factor of the absorbing gas. Defining  the 
absorption depth as $\Delta S(v) = S_{\rm cont}-S(v)$, we can rewrite this as
\begin{equation}
\Delta S(v) = S_{\rm cont} \, c_f \left(1- e^{-\tau(v)}\right).\label{eq:deltaSexp}
\end{equation}
In the optically thin limit, this becomes
\begin{equation}
\tau(v) \simeq \frac{1}{c_f}\frac{\Delta S(v)}{S_{\rm cont}}\label{eq:deltaS}.
\end{equation}
The \hi column density in \cm is related to the optical-depth profile through
\begin{equation}
N_{\rm \ion{H}{i}} = 1.823 \times 10^{18} \frac{T_s}{c_f}
\int \tau(v)\,dv. \label{eq:nhicnm}
\end{equation}
Here, with $v$ in \kms and $T_s$ in K, the numerical factor converts from K\,\kms to \cm.
This expression shows that, once the optical depth spectrum $\tau(v)$ has been determined, the inferred column density scales linearly with the assumed spin temperature $T_s$ and inversely with the covering factor $c_f$.
The latter is not always known, and often one assumes $c_f = 1$.


In the optically thin limit, the \hi column density is related to the brightness temperature spectrum $T_B(v)$ through Eqs.\ \ref{eq:smalltau} and \ref{eq:nhicnm}.
It can
be written as
\begin{equation}
N_{\rm \ion{H}{i}}
= 1.823 \times 10^{18} \int T_B(v)\, dv\label{eq:nhiwnm}.
\end{equation}
This expression is independent of optical depth and shows that optically thin gas can contribute substantially to the observed emission while producing only weak absorption.
This difference underlies the fundamental distinction between absorption and emission:
absorption traces the CNM through $\tau(v)$,
whereas emission traces the combined contribution of both CNM and WNM  through $T_B(v)$.

We model the absorption and emission spectra as combinations of Gaussian components. Since the CNM produces measurable absorption, we parameterize it directly through its optical-depth profile, $\tau(v)$. The corresponding CNM emission $T_B(v)$
can then be derived from $\tau(v)$ and the radiative transfer equation (Eq.~\ref{eq:tau}).

For the absorption spectrum, which in practice traces the CNM-dominated absorbing gas, the optical depth can be
written as the sum of $N$ Gaussian components, so that
\begin{equation}
\tau(v) = \sum_{n=0}^{N-1} \tau_{0,n}
\exp[-(v - v_{0,n})^2/(2\sigma_{v,n}^2)].\label{eq:ncomp}
\end{equation}
Each Gaussian represents a different CNM cloud or component,
characterised by its peak optical depth $\tau_{0,n}$, central velocity
$v_{0,n}$, and velocity dispersion $\sigma_{v,n}$. 
The full widths at half maximum are given by $\mathrm{FWHM} = 2\sqrt{2\ln2}\,\sigma_v \approx 2.35\sigma_v$.

The emission spectrum, in contrast, contains contributions from both
the CNM and the WNM. We write 
its expected brightness temperature as
\begin{equation}
T_{\rm exp}(v) = T_{B,{\rm CNM}}(v) + T_{B,{\rm WNM}}(v).
\end{equation}
For the CNM, the emission follows directly from Eq.~\ref{eq:tau}.
In contrast, the WNM has low optical depth and does not typically produce detectable absorption (but see \citealt{killerby2025,patra2018}). It must therefore be constrained from the emission spectrum alone.
We model the WNM directly in brightness temperature space, describing its contribution as a sum of $K$ Gaussian components,
\begin{equation}
T_{B,{\rm WNM}}(v) =
\sum_{k=0}^{K-1}
T_{0,k}
\exp[-(v - v_{0,k})^2/(2\sigma_{v,k}^2)].
\end{equation}
These WNM spectra must be modified to account for absorption by CNM components present between us and the WNM components along the line of sight.
For the case of a single CNM component, we introduce a factor $F_k$ that describes the fraction of WNM emission lying in front of the absorbing CNM along the line of sight.
This yields
\begin{equation}
T_{B,{\rm WNM}}(v) =
\sum_{k=0}^{K-1}
[ F_k + (1 - F_k)e^{-\tau(v)}]\,
T_{0,k}
\exp\left[-\frac{(v - v_{0,k})^2}{2\sigma_{v,k}^2}\right].\label{eq:wnm}
\end{equation}
In general, $F_k$ is not used as a fit parameter. Instead we assume a number of values $F_k = 0$, $0.5$, and $1$ to cover the possible geometries. These correspond to the WNM lying entirely
behind the CNM (maximum absorption), partially mixed with the CNM (intermediate absorption), or entirely in front of the CNM (no absorption), respectively. 

When more than one CNM component is present, the radiative transfer becomes more complex, as each component emits its own radiation while also absorbing radiation from components located behind it along the line of sight.
The total expected spectrum
can be written as
\begin{equation}
\begin{aligned}
T_{\rm exp}(v) &= 
\sum_{n=0}^{N-1} T_{s,n}(1-e^{-\tau_n(v)})
\exp\!\left[-\sum_{m<n}\tau_m(v)\right] \\
&+ \sum_{k=0}^{K-1}
\left[ F_k + (1-F_k)\exp\left(-\sum_{n=0}^{N-1}\tau_n(v)\right) \right]
T_{0,k} \\
&\times \exp\left[-\frac{(v-v_{0,k})^2}{2\sigma_{v,k}^2}\right].
\end{aligned}
\end{equation}
The observed emission spectrum alone does not uniquely determine the relative
ordering of CNM and WNM components along the line of sight, as multiple
arrangements can produce similar spectra. To address this degeneracy, we
explore all permutations of the CNM-WNM geometry. For a single CNM component and
$K$ WNM components, this results in $3^K$ distinct configurations, corresponding
to the three possible values of $F_k$. In the general case of $N$ CNM components,
the number of possible permutations scales as $N!\,3^K$, reflecting both the
ordering of the CNM clouds along the line of sight and the placement of the WNM
components relative to them. In practice, the number of distinct solutions is
somewhat smaller, as permutations involving nearly identical components produce spectra that are observationally indistinguishable.

In practice, the optical-depth spectrum, derived from the absorption spectrum, is first
fitted to determine the number of CNM components $N$. This fixes the CNM
optical-depth parameters $\tau_{0,n}$, the central velocities $v_{0,n}$, and the
velocity widths of the individual components $\sigma_{v,n}$. The observed
emission spectrum is then fitted to determine the number of WNM components $K$, where
the CNM optical-depth profiles $\tau_n(v)$ are treated as fixed parameters, and the CNM enters the emission modeling only through the spin
temperatures $T_{s,n}$, which are free parameters.

Even with $N$ and $K$ fixed, an ambiguity remains in the relative line-of-sight
arrangement of the CNM and WNM components, as encoded in the parameters $F_k$.
Following the terminology of \citet{heiles2003a}, each distinct combination of
$F_k$ values for fixed $N$ and $K$ defines a ``trial''. Each trial is fitted to
the observed emission spectrum. Because the data do not generally uniquely
determine the true CNM-WNM geometry, no single trial can be preferred a priori.
Instead, physical parameters are derived by averaging over all acceptable
trials, that is, over all possible (combinations of) choices of $F_k$ for fixed $N$ and $K$. In
particular, the final spin temperature $T_s$ is computed as a weighted average
over all trials (for a given $N$ and $K$), with weights determined by the goodness of fit, following
Equations~(21a) and~(21b) of \citet{heiles2003a}.

From the fitted components, several additional physical diagnostics are
computed. The maximum kinetic temperature of each component is given
by
$T_{k,{\rm max}} = m_H/(8 k_B \ln 2)\,  \Delta v_{\rm FWHM}^2
= 21.866 \, \Delta v_{\rm FWHM}^2$,
providing an upper limit on the true kinetic temperature in the absence of non-thermal broadening. This is useful because in a CNM--WNM mixture the spin temperature generally differs from the kinetic temperature.

The CNM and WNM column densities are given by Eq.~\ref{eq:nhicnm} and Eq.~\ref{eq:nhiwnm}, respectively (with $c_f=1$). 
From these, the CNM fraction is defined as
\begin{equation}
f_{\rm CNM} =
\frac{N_{\rm \ion{H}{i},CNM}}
     {N_{\rm \ion{H}{i},CNM} + N_{\rm \ion{H}{i},WNM}},
\end{equation}
which quantifies the relative contribution of cold gas to the total
neutral hydrogen column density.

Finally, we compute the density-weighted mean spin temperature
\citep{dickey2000},
\begin{equation}
\langle T_s \rangle =
\frac{\int T_B(v)\,dv}
     {\int [1 - e^{-\tau(v)}]\,dv}.\label{eq:meanspin}
\end{equation}
This quantity provides a single effective temperature that reflects the relative importance of absorption and emission, with the denominator dominated by contributions from gas with significant optical depth (primarily the CNM)
and the numerator incorporating emission from both CNM and WNM. We can understand this as follows: starting from Eq.~\ref{eq:tau}, we get
$T_s(v) = T_B(v)/[1 - e^{-\tau(v)}]$, but this quantity is poorly
defined when $\tau(v)$ is small (as will be the case for the WNM). Instead, we can define a single effective temperature $\langle T_s \rangle$ that relates the velocity-integrated brightness temperatures and optical depth: 
$\int T_B(v)\,dv = \langle T_s \rangle \int [1 - e^{-\tau(v)}]\,dv$,
resulting in the above expression for $\langle T_s \rangle$.

Finally, we emphasise an important limitation of combined emission–absorption analyses. Absorption probes a narrow line of sight toward a compact background source, whereas the corresponding emission spectrum is obtained from nearby positions and averaged over the larger telescope beam. This difference in spatial sampling becomes increasingly important at larger distances, where the physical scale of the beam grows and variations in the \HI distribution and kinematics become more significant. The resulting mismatch introduces a systematic uncertainty that must be accounted for in the modeling and interpretation.

\section{Results}
\subsection{Continuum source selection}

In searching for absorption in the MHONGOOSE galaxies, we adopt a lower peak flux limit of $S_{\rm cont} = 0.7$\,\mJybeam for the continuum sources considered.
 This threshold is set by the achievable optical-depth sensitivity of the data. For small optical depths, the 1$\sigma$ uncertainty in optical depth is given by
$\sigma_\tau = {\sigma_{\rm I}}/{S_{\rm cont}}$,
where $\sigma_{\rm I}$ is the per-channel rms noise. For the \rzero data, $\sigma_{\rm I} = 0.22$ \mJybeam. For $S_{\rm cont} = 0.7$ \mJybeam, this yields a 1$\sigma$ optical-depth sensitivity of $\sigma_\tau \simeq 0.3$.

Requiring a $3\sigma$ detection in optical depth implies sensitivity to peak optical depths of $\tau_0 \sim  0.9$. 
At this limit, we are sensitive to absorption features with peak optical depths of order unity per channel.
Sources fainter than 0.7\,\mJybeam would only allow the detection of absorption with very high optical depth. We exclude such sources from further consideration. 


We use the Python Blob Detection and Source Finder  (\texttt{pybdsf}; \citealt{mohan2015}) to 
identify all continuum sources in the primary-beam-corrected continuum images with peak flux densities $>0.7$\,\mJybeam that are spatially coincident with the \hi disks, as defined by the \rzero zeroth-moment maps.
These maps have an average $3\sigma$ column density sensitivity of $6 \times 10^{19}$ \cm.
The 0.7 \mJybeam peak flux represents a signal-to-noise of $\sim 200$ in the continuum images, 
so the source detection is unambiguous.
 
In the centers of galaxies NGC 1566 (J0419--54) and NGC 1672 (J0445--59), we detect absorption lines with widths of
${\sim}60-100$\,\kms associated with the AGN in these galaxies. We do not study these here, and exclude these central sources from the catalog. 
The final list contains 56 continuum sources distributed over 21 galaxies. 
Nine MHONGOOSE galaxies were not included, as their \HI distributions do not overlap with any sufficiently bright continuum sources.


\subsection{Detections and non-detections}

\begin{figure}[t]
\centering
\includegraphics[width=0.9\columnwidth]{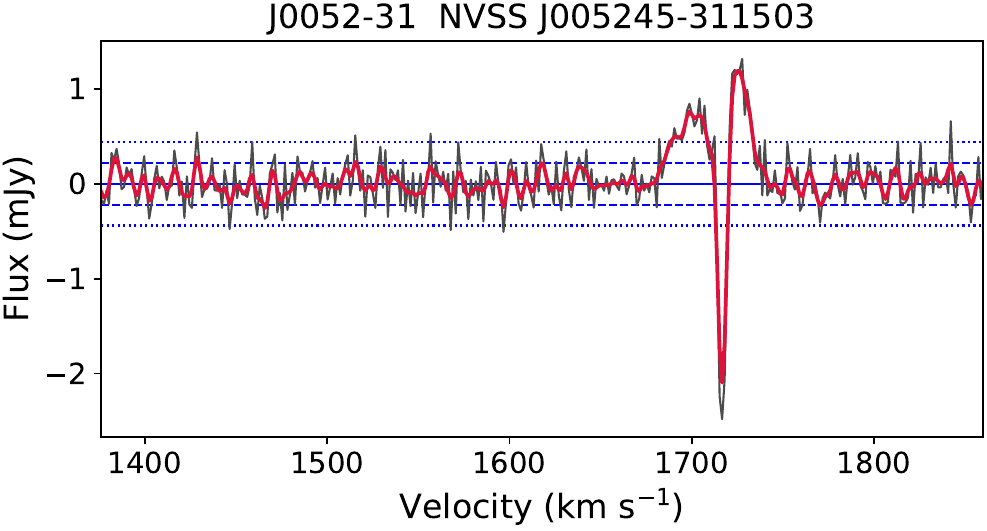}\\
\vspace{4pt}
\includegraphics[width=0.9\columnwidth]{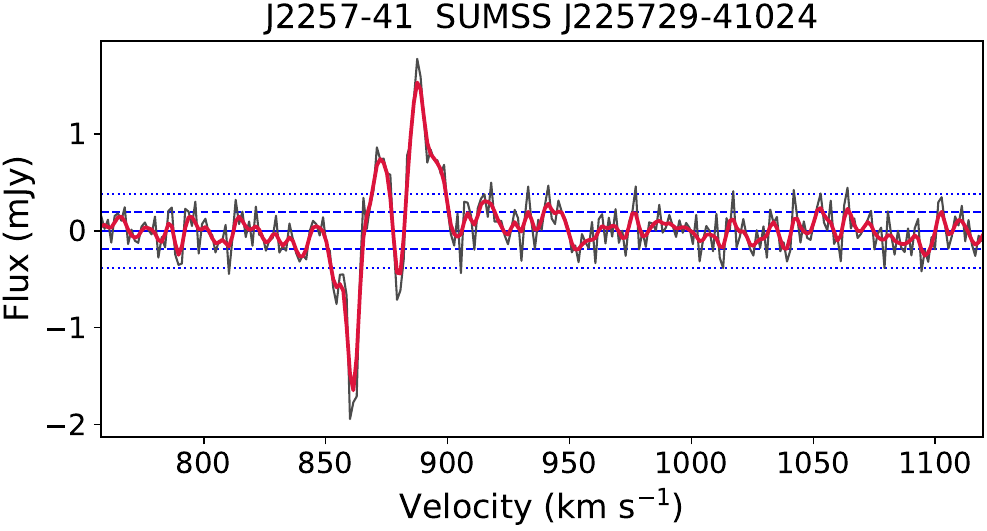}\\
\vspace{4pt}
\includegraphics[width=0.9\columnwidth]{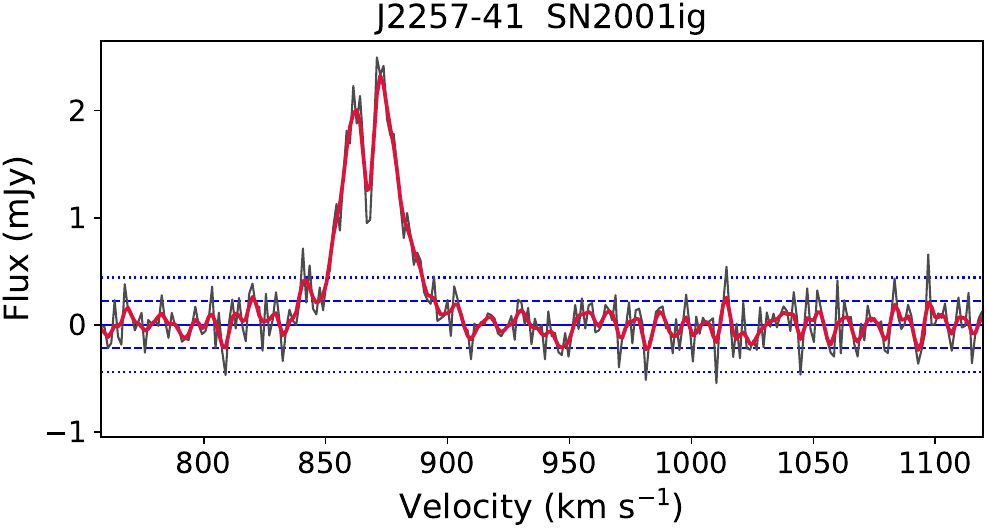}
\caption{Spectra of the three absorption profiles. From top to bottom: NVSS~J005245$-$311503 in NGC~289 (J0052$-$31), SUMSS~J225729$-$410241 in NGC~7424 (J2257$-$41), and SN2001ig also in NGC~7424 (J2257$-$41). The original spectra are shown in black, with a five-point Hanning-smoothed version overplotted in red. The 0, $\pm 1\sigma$, and $\pm 2\sigma$ levels of the unsmoothed spectrum are indicated by solid, dashed, and dotted lines, respectively.}

\label{fig:detectionspectra}
\end{figure}

We extract spectra from the \rzero data cubes at the positions of the continuum sources and detect absorption toward three of them. 
One detection is found in NGC~289 (J0052--31), toward NVSS J005245--311503 (NRAO VLA Sky Survey; \citealt{condon1998}). 
Two additional detections in close proximity are found in NGC~7424 (J2257--41), toward SUMSS J225729--410241 (Sydney University Molonglo Sky Survey; \citealt{mauch2003}) and the supernova SN2001ig.
These spectra are shown in Fig.~\ref{fig:detectionspectra}. We will discuss them in more detail in Sect.~\ref{sec:spectra}. The properties of the three continuum sources are listed in Table~\ref{tab:contcat}.

\begin{table*}
\small
\centering
\caption{Continuum sources with peak fluxes $> 4$ mJy overlapping the \hi disks}
\begin{tabular}{l l c c @{\hspace{-5pt}} r r l  r}
\hline
\hline
\noalign{\vskip 2pt}
Galaxy & Radio ID  & $\alpha\,(2000.0)$ & $\delta\,(2000.0)$ & $S_{\rm peak}$ & $F$ & res. & $N_{\rm \ion{H}{i}}$ \\
 & & $(^h\,^m\,^s)$ & $(\degree\,'\,'')$ & (mJy bm$^{-1}$) & (mJy) & & ($\times 10^{20}$ cm$^{-2}$)\\
(1) &  (2) &  (3) &  (4) & (5) & (6) & (7) & (8) \\ 
\hline
\noalign{\vskip 2pt}
\multicolumn{2}{l}{\bf Detections:} &&&&&&\\
J0052--31 & NVSS J005245--311503  & 00 52 45.622 & --31 15 03.371 & 18.3 & 18.4 & N & 11.77 $\pm$   0.46 \\  
J2257--41 & SUMSS J225729--410241 & 22 57 29.631 & --41 02 39.593  & 45.9 & 47.2 & N & 12.00 $\pm$   0.49\\ 
J2257--41 & SN2001ig & 22 57 30.731 & --41 02 26.141 & 10.7 & 10.8& N & 1.48 $\pm$   0.35\\
\hline
\noalign{\vskip 2pt}
\multicolumn{2}{l}{\bf Non-detections:} &&&&&&\\
J0135--41 & central star burst region    & 01 35 06.983 & --41 26 11.986 & 5.6 & 7.6 & Y* & 32.13 $\pm$   0.60 \\
J0309--41 & SUMSS J030941--410006 & 03 09 41.412 & --41 00 06.281 & 59.4 & 60.6 & N & 3.71 $\pm$   0.43\\ 
J0335--24 & NVSS J033451--245721  & 03 34 51.368 & --24 57 14.402  & 5.3 & 5.6 & N & 6.36 $\pm$   0.41\\
J0419--54 & no previous ID & 04 19 55.822  & --55 01 36.696 &    5.3  & 5.5 & N & 3.48 $\pm$   0.40\\
J0419--54 & SUMSS J042007--545143(S)\tablefootmark{a} & 04 20 05.025 & --54 52 10.580 & 6.4 & 8.1 & Y & 4.00 $\pm$   0.45\\
J0419--54 & SUMSS J042007--545143(N)\tablefootmark{a} & 04 20 08.522 & --54 51 32.789  & 14.3 & 15.6 & Y & 0.62 $\pm$   0.35\\ 
J0419--54 & SUMSS J042015--545345 & 04 20 15.307 & --54 53 46.068 & 30.3 &  40.7 & N & 1.57 $\pm$   0.45\\ 
J0429--27 & NVSS J042936--272427 &  04 29 36.491 & --27 24 26.191 & 4.9 & 4.9 & N &1.07 $\pm$   0.38 \\
J0454--53 & SUMSS J045418--531841 & 04 54 18.228 & --53 18 43.362 &  6.0 & 6.1 & N & 0.85 $\pm$   0.33\\
J1303--17b& NVSS J130311--172314 & 13 03 11.249 & --17 23 14.612 &  4.6 &  5.0 &  N &1.11 $\pm$   0.32 \\ 
J1337--28 & NVSS J133718--280405  & 13 37 18.604 & --28 04 03.938 & 8.0 & 8.9 & N & 1.92 $\pm$   0.34\\
J2257--41 & SUMSS J225639--410425 & 22 56 40.421 & --41 04 32.531 & 4.6 & 4.6 & N & 1.64 $\pm$   0.39\\
J2257--41 & no previous ID & 22 57 09.401 & --40 59 35.264 & 4.0 & 4.0 & N & 13.86 $\pm$   0.49\\
J2357--32 & NVSS J235730--323728 &  23 57 30.530  & --32 37 28.718  & 6.7 & 6.8 & N & 5.32 $\pm$   0.46\\
\hline
\end{tabular}
\tablefoot{(1) HIPASS name of foreground galaxy (2) Identification in radio catalog (3) Right Ascension (J2000.0) (4) Declination (J2000.0) (5) Peak flux (6) Total flux (7) Resolved source? * indicates association with H$\alpha$ (8) Foreground \hi column density\\
\tablefoottext{a}{This source is a radio galaxy with two lobes. (N) and (S) refer to the northern and southern lobes respectively.}}
\label{tab:contcat}
\end{table*}

The remaining 53 spectra show no obvious evidence for absorption. We quantify this by using a simple detection criterion in which the spectra are smoothed with a five-point Hanning filter. For a detection, absorption is required to extend over more than two original channel widths below $-3\sigma$ in the smoothed spectrum, and to be located within the velocity range of the galaxy as measured from the first-moment maps. 
A five-point Hanning smoothing changes the effective spectral resolution to approximately twice the original channel width, and the requirement that absorption extends over more than two original channels ensures that we are sensitive to features broader than a single smoothed resolution element.

To illustrate, we list the properties of the brightest non-detection background sources in Table~\ref{tab:contcat} and show the corresponding spectra in Fig.~\ref{fig:nodetections}. 
The difference between the three detections and the non-detections is clear. 
All detections are highly significant, with peak S/N values $>10$, and no detections are found at lower peak S/N.

\begin{figure}[t]
\centering
\includegraphics[width=0.99\columnwidth]{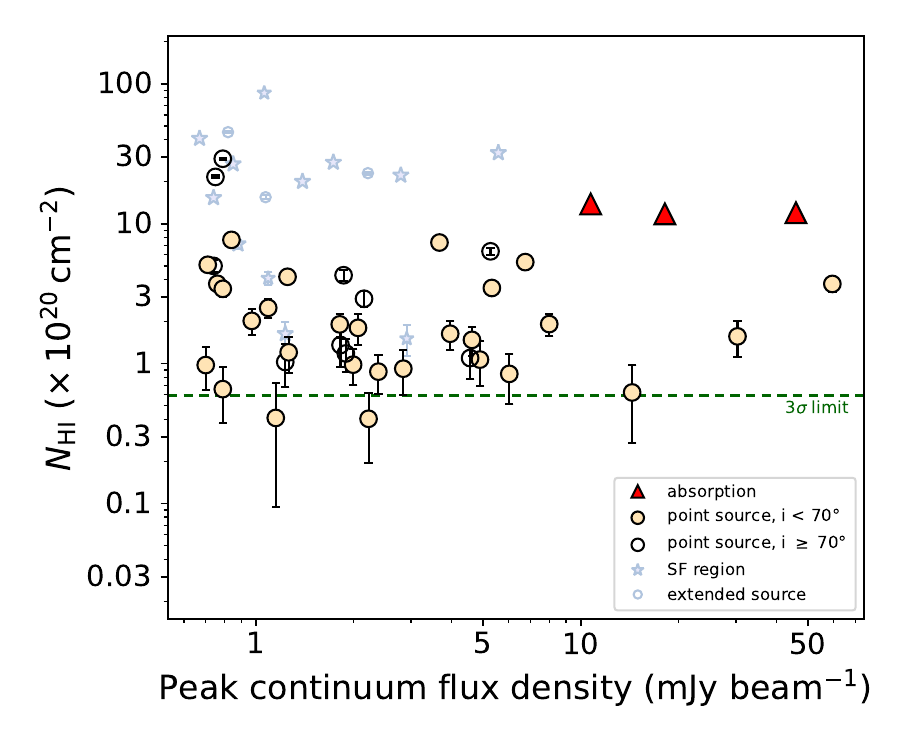}
\caption{
\hi emission column densities at the positions of the continuum sources plotted against continuum peak flux. The column densities are not corrected for inclination.
Red triangles indicate detections. Circles indicate non-detections toward unresolved, non-H$\alpha$ sources, with filled symbols for $i<70\degree$ and open symbols for $i\geq 70\degree$. Small stars indicate H$\alpha$ star-forming regions, and small circles resolved sources not associated with H$\alpha$. 
Filled circles mark the ``clean'' sample of unresolved, non-H$\alpha$ sources in low-inclination galaxies. 
The horizontal dashed line shows the $3\sigma$ (16 \kms) column density limit of $\log(N_{\rm \ion{H}{i}}/{\rm cm}^{-2}) = 19.77$.}
\label{fig:peakNHI}
\end{figure}

To quantify the presence or absence of absorption as a function of the strength of the background continuum source and of the foreground \hi emission,  we measure the \hi emission column density values in the \rzero zeroth-moment maps at the positions of the non-detection continuum sources. For the sources with absorption, we measure the column density in a one-beam-wide annulus with inner radius one beam, centred on the continuum position.
We plot these column densities against the peak flux of the continuum sources in Fig.\ \ref{fig:peakNHI}.
We detect \HI absorption only where both the continuum peak flux and the emission column density are high.
The three absorption spectra all occur at emission column densities ${\sim}1 \times 10^{21}$ \cm. Note the difference between the two strongest continuum sources. The ${\sim}40$ \mJybeam source SUMSS J225729--410241 shows absorption with a  ${\sim}10^{21}$ \cm foreground column density.
The stronger (${\sim}60$\,\mJybeam) source SUMSS J030941--410006 has a foreground column density only a factor of ${\sim}3$ lower, yet shows no absorption.
Sources fainter than 9 \mJybeam show no absorption regardless of column density. 

The lack of absorption despite high column densities could, in some cases, be caused by projection effects along the line of sight due to high inclination values. 
This can allow lower column density \HI (primarily WNM) to accumulate along the line of sight, mimicking a high column density without a significant CNM component.
In Fig.~\ref{fig:peakNHI} we indicate all $i > 70\degree$ galaxies with a different symbol, where inclinations are taken from \citet{deblok2024}. Of the 18 galaxies with low-$S_{\rm peak}$ and high-$N_{\rm HI}$ sight-lines, eight have a high inclination, but ten do not. For the latter, projection effects cannot cause the high column densities.

A fraction of the continuum sources turns out to be (somewhat) extended. 
This can reduce the probability of detecting absorption, as the effective covering factor $c_f$ may be less than unity.
We identified sources with major axis size $> 1.1\, b_{\rm max}$ or minor axis size $> 1.1\, b_{\rm min}$, where $b_{\rm max}$ and $b_{\rm min}$ are the major and minor axis of the beam, respectively.
Comparison with H$\alpha$ imaging from the Survey for Ionization in Neutral Gas Galaxies  (SINGG; \citealt{meurer2006}) shows that most of these are associated with star-forming regions in the target galaxies. Of the 56 sources, 44 are unresolved (three with H$\alpha$) and 12 resolved (nine with H$\alpha$).
Selecting unresolved, non-H$\alpha$ sources behind galaxies with $i<70\degree$ yields 31 sources in 14 galaxies. The different source types are indicated in Fig.~\ref{fig:peakNHI}. The ``clean'' sample contains no sources with both low $S_{\rm peak}$ and high $N_{\rm HI}$ sight lines. We return to this in Sect.~\ref{sec:abslimits}.

\section{Absorption spectra\label{sec:spectra}}

Here we present and analyse the three absorption spectra in more detail. We discuss the properties of the foreground galaxies, derive the intrinsic absorption-only spectra and discuss the additional difficulties of deriving and interpreting the emission spectra in objects at these distances. We present the properties of the CNM and WNM components along these three lines of sight.

\subsection{Absorption in NGC 289}
\subsubsection{Properties of the galaxy and continuum source}
We detect a single absorption line of sight in NGC 289 (J0052--31).
This galaxy is among the most massive galaxies in the MHONGOOSE sample. 
It hosts a bright inner disk with a stellar bar and a faint outer low-surface-brightness disk containing several star-forming spiral arms.
NGC~289 hosts a low-luminosity AGN, and is usually classified as a Seyfert galaxy. 
The galaxy is gas-rich, with its \HI mass ($2.2 \times 10^{10}$ \msun) almost equal to  its stellar mass ($2.7 \times 10^{10}$ \msun).
We assume a distance of 21.5 Mpc \citep{leroy2019}, corresponding to $1'' = 0.1$ kpc. NGC~289 is not part of any major group or cluster, but resides in a low-density environment. \citet{kourkchi2017} list it as the main galaxy in its own sparse low-density association. 

NGC~289 was studied in \HI before by \citet{walsh1997} using the Australia Telescope Compact Array (ATCA). In addition to the properties listed above, they also 
note the flat and extended rotation curve (with a maximum radius of over 50 kpc, adjusted to our assumed distance), the dark matter dominance of the low-surface brightness disk and the significant recent star formation in the far outer disk. 

\begin{figure}[t]
\centering
\includegraphics[width=0.99\columnwidth]{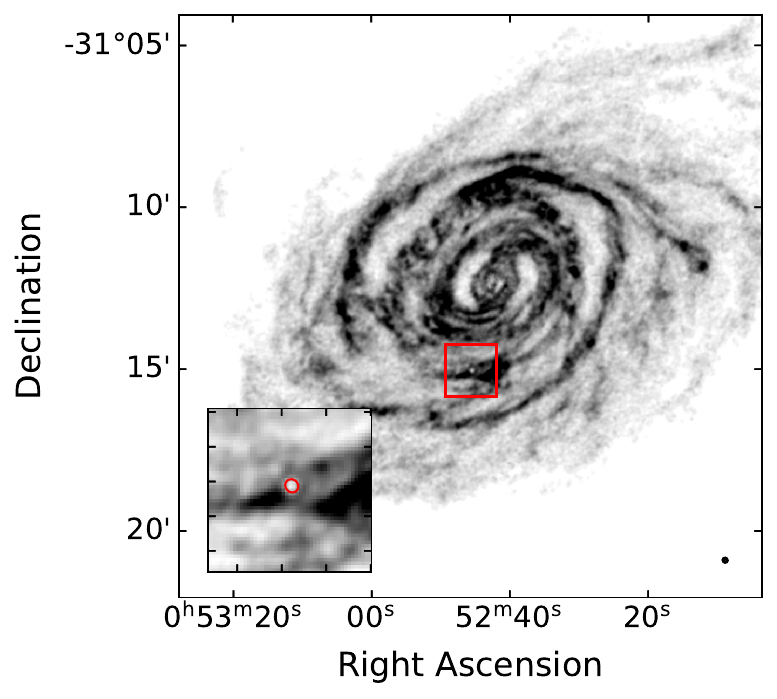}

\caption{The \rzero\ zeroth-moment map of NGC~289. The red box in the main panel ($1.6' \times 1.6'$) marks the position of the absorption feature. The bottom-left inset shows a zoom of this region. The red ellipse indicates the continuum source. The \rzero\ beam is shown in the lower right.}
\label{fig:n289mom0}
\end{figure}

Inspection of the \rzero \hi zeroth-moment map already reveals the presence of the absorption feature in the outer \HI disk as shown in Figure \ref{fig:n289mom0}. 
It is also visible in the {\tt r05\_t00} data (with an 
angular resolution of $11''$), but not in the {\tt r10\_t00} data (resolution $22''$) or at lower resolutions, where the larger beam and increased emission filling dilute the absorption feature.

A Gaussian fit to the source in the continuum image shows it is unresolved.
The flux listed for this source in NVSS at 1.4 GHz is $20.4 \pm 1.1$ mJy, consistent with the MHONGOOSE value listed in Tab.~\ref{tab:contcat}.
NVSS J005245--311503 has an alternative identification (WISEA 
J005245.58--311503.2), and is visible in WISE bands W1 
(\SI{3.4}{\micro\metre}) and W2 (\SI{4.6}{\micro\metre}).
It is not visible in W3 and 
W4 (12 and \SI{22}{\micro\metre}), respectively.
In the Legacy Survey \citep{dey2019} we also find a faint optical counterpart at the position of the 
infrared source. Its colour is much redder than the surrounding disk emission 
of NGC 289, indicating it is likely a background source seen through the galactic disk.
The projected galactocentric distance (impact parameter) of this detection is ${\sim}17.5$ kpc. 

\subsubsection{The absorption spectrum\label{sec:absorption}}

\erwin{Both the WNM and the CNM can cause absorption, though Eqs.~\ref{eq:nhicnm} and \ref{eq:nhiwnm} indicate that any absorption signal will be dominated by that of CNM absorption. Detections of WNM absorption in the MW exist, but require extremely sensitive observations (see, e.g., \citealt{killerby2025} and \citealt{patra2018}). Certainly, at the sensitivities of our observations, we expect the absorption signal to be fully dominated by the CNM.}
    
\erwin{Emission profiles, in contrast, contain of a more equal mix of CNM and WNM contributions.} For a pure absorption measurement it is important to image the CNM absorption in isolation and avoid any contamination from the CNM and WNM emission.  We created cubes containing only the absorption. As the absorption is due to a point source, while the emission is extended, this can be achieved by suppressing the signal on the short baselines (e.g., \citealt{dickey2003}).  We found that a robust value of $-1$ combined with removing (flagging) all baselines with a length below 1000\,m, gives a good compromise between removing the \HI emission and not amplifying the noise significantly. As we are interested in the absorption with respect to the continuum background, the absorption cube was created using measurement sets where the continuum was not subtracted. The beam size of this cube is $5.27''\, \times\, 5.10''$ with $1''$ pixels. 

\begin{figure}[t]
\centering
\includegraphics[width=0.9\columnwidth]{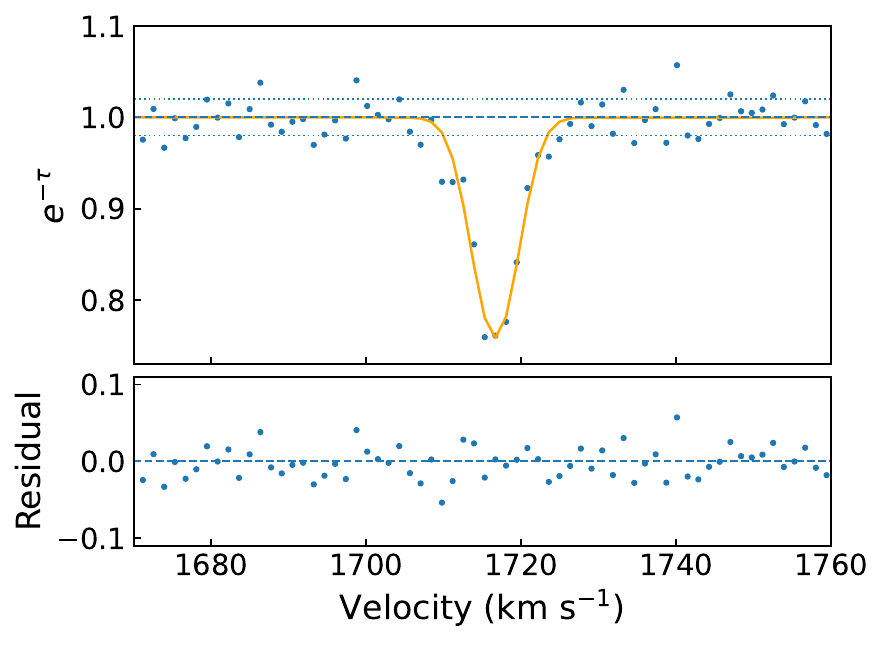}
\caption{Top: intrinsic absorption spectrum of NVSS~J005245$-$311503 in NGC~289 derived from the absorption-only data cube, shown as optical depth $\tau$. The dashed line indicates zero optical depth and the dotted lines the $\pm 1\sigma$ levels. A single-component Gaussian fit is overplotted. Bottom: residuals with respect to the Gaussian fit.}

\label{fig:N289absorption}
\end{figure}

Due to the removal of the short baselines and the low robustness factor, the noise in the cube increases by about a factor of two to 0.41\,\mJybeam. This loss of sensitivity is also why these cubes are not used for source and absorption detection ab initio.
To measure the intrinsic CNM absorption spectrum, we consider only pixels within the FWHM of the synthesized beam centred on the background continuum source.
For each of these pixels, we measure the continuum flux density $S_{\rm cont}$ from line-free channels in the velocity range 1300–1600~\kms.
Following \citet{dickey1992}, we then construct an averaged absorption spectrum from these pixel spectra. For each pixel, we use the spectrum $S(v)$ from data without continuum subtraction and form the normalized spectrum $S(v)/S_{\rm cont}$, corresponding to $e^{-\tau(v)}$ for $c_f=1$ (Eq.~\ref{eq:deltaSexp}).
These normalized spectra are combined using weights proportional to the square of the continuum flux density in each pixel. This weighting emphasizes high signal-to-noise pixels while preserving the intrinsic absorption profile. The resulting weighted average yields the intrinsic absorption spectrum in units of $e^{-\tau(v)}$. The final spectrum is shown in Fig.~\ref{fig:N289absorption}.



\subsubsection{The emission spectrum}

Absorption samples a pencil-beam line of sight toward a background continuum source. The corresponding emission spectrum, however, cannot be measured at that exact position and is therefore derived from nearby regions using a larger beam.
Any combined absorption–emission analysis must 
assume that properties of the ISM at the absorption and emission positions are comparable.  This assumption can often be justified for MW studies
where the physical resolution can reach sub-parsec scales.

The situation quickly becomes more complicated with increasing distance. 
\citet{chen2025} note in their study of the LMC and the SMC 
that with a spatial resolution of $\sim 8$ pc, an offset of a beam or two may already 
lead to emission spectra that are not tracing exactly the same regions as the absorption spectrum, and contain different emission components.

In many previous studies the emission spectrum is derived by interpolating across the absorber position 
(e.g., \citealt{dickey1992}, or by averaging the spectra in a region around the absorber. For example, in their study of NGC 6822,  
\citet{pingel2024} use an annulus with an inner diameter of two beam widths and an outer 
diameter of four beam widths (i.e., with a thickness of one beam width). At the distance of NGC 6822 this annulus spans $\sim 60$ pc.
For NGC~289 and NGC~7424, our typical beam size measures $\sim 0.8$ kpc and $\sim 0.3$ kpc, respectively, i.e., an order of magnitude larger. This implies that opposite ends of such an annulus are separated by more than a kpc. 

\begin{figure}[t]
\centering
\includegraphics[width=0.9\columnwidth]{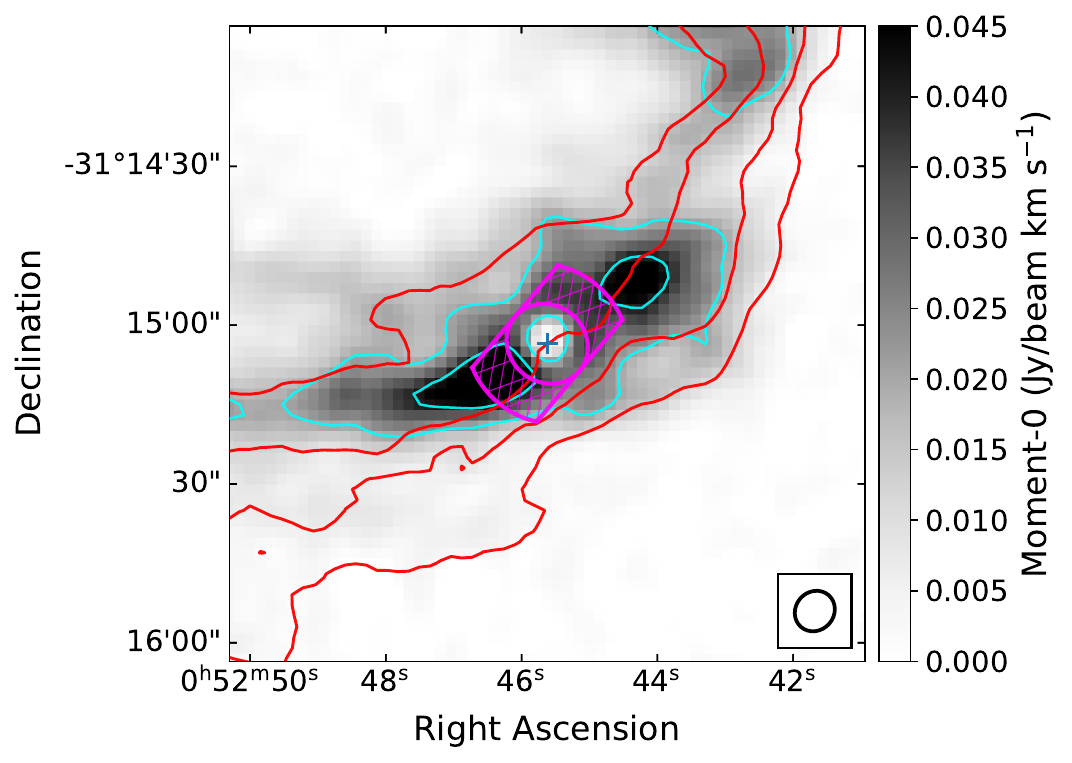}
\caption{Zeroth-moment map of the region around the NGC~289 absorber NVSS~J005245$-$311503 in grayscale. The map was created by collapsing channels of the \rzero\ data cube between 1708 and 1725 \kms, corresponding to $\pm 3\sigma$ around the CNM central velocity. No masking was applied. Cyan contours show the \hi\ column density at 0.02 and 0.04 Jy beam$^{-1}$ \kms\ (corresponding to $4.4 \times 10^{20}$ and $8.8 \times 10^{20}$ \cm, respectively). Red contours show velocities of 1692 (southernmost), 1702, 1712, and 1722 \kms\ (northernmost). The pink hatched regions indicate the intersection of the annulus with lines at ${\rm PA} = -55\degree$, used to derive the emission spectrum. The \rzero\ beam is shown in the lower right.}

\label{fig:N289emissionmom0}
\end{figure}

This complication can be seen in the significant variation in \HI column density in the area 
immediately surrounding the absorption in NGC 289. Figure \ref{fig:N289emissionmom0} shows a zeroth-moment map of the emission around the absorber, created by summing the channels in a narrow range in velocity around the central absorption velocity (cf.\ Fig.~\ref{fig:N289absorption}). Also shown are a number of iso-velocity contours indicating the kinematics in this region.

Inspection of the figure shows that the absorber lies behind an \HI over-density that crosses 
it in a roughly south-east to north-west direction. This over-density is part of a larger spiral arm 
(Fig.~\ref{fig:n289mom0}). The iso-velocity contours run approximately parallel with the filament.
Between opposite ends of the absorption feature there is a velocity gradient of $\sim 15$ \kms.
Velocities at these positions are not representative of those at the absorber location.
Averaging in, for example, an 
annulus would result in an artificial broadening of the spectrum. In principle, one could correct for this by removing the rotational signal, and aligning the spectra at the velocity of the absorber, but this would introduce additional assumptions.

We define a region that follows the observed \hi and velocity distribution rather than a purely geometrical model.
We select two regions on either side of the absorber, aligned with the filament and velocity contours.
These regions are defined as the intersection of an annulus (inner radius one beam, outer radius two beams) with two parallel lines on either side of the absorber at a position angle of $-55\degree$, each offset by one beam.
These regions are  indicated in 
Fig.~\ref{fig:N289emissionmom0}. 
The position angle approximates that of the filament and velocity contours.

\begin{figure}[t]
\centering
\includegraphics[width=0.99\columnwidth]{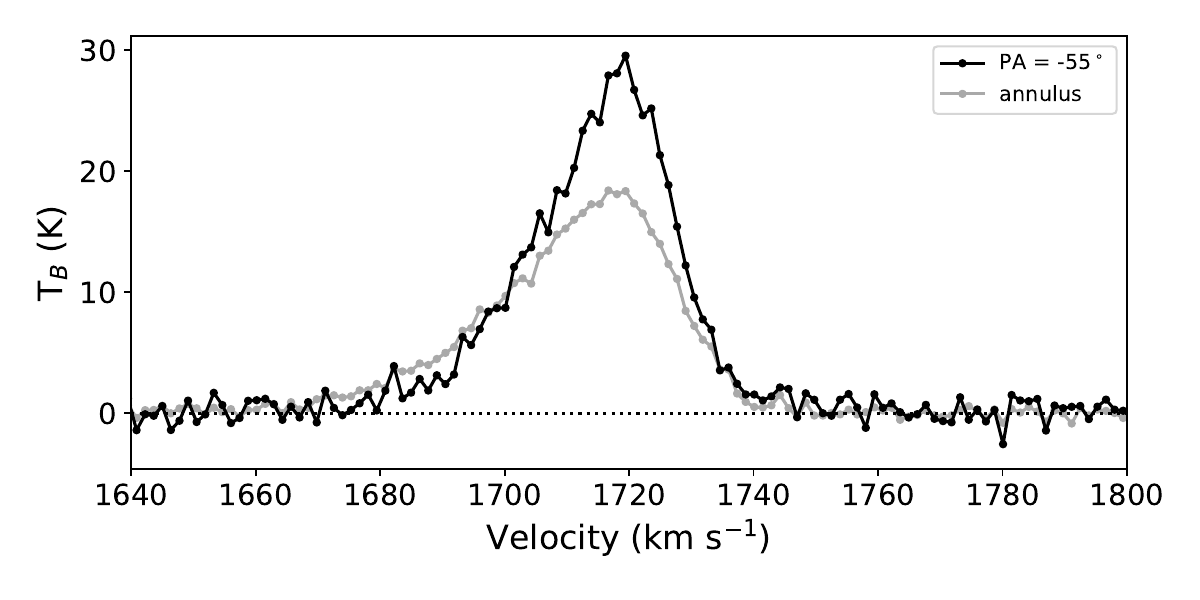}
\caption{Variation in emission spectra around the absorber NVSS~J005245$-$311503. The gray line shows the average spectrum derived from an annulus with inner radius equal to one beam width and outer radius of two beam widths. The black line shows the spectrum derived from the intersection of this annulus with lines at a position angle of $-55\degree$. This region is shown in Fig.~\ref{fig:N289absorption}. Note the low-velocity wing in the annulus spectrum, as well as the difference in peak brightness temperature.}

\label{fig:emissionvariation}
\end{figure}

In Fig.\ \ref{fig:emissionvariation} we show the emission spectrum for these intersection regions as well as that from the full annulus. 
The difference between full and partial annulus profiles is obvious, and emphasizes both the care required in deriving emission spectra and the associated systematic uncertainties.

\subsubsection{Analysis of the spectra\label{sec:analysis}}
We first derive the parameters of the absorption spectrum, which is well described by a single Gaussian (as shown in Fig.~\ref{fig:N289absorption}).
We obtain a peak optical depth of $\tau_0 = 0.27 \pm  0.01$, a 
velocity dispersion of $\sigma_0 = 2.90 \pm 0.16$\,\kms (FWHM$ =  6.83 \pm 0.38$\,\kms) and a central 
velocity of 
$v_0 = 1716.7 \pm 0.16$\,\kms. 
These, and other fit parameters discussed later, are listed in Table~\ref{tab:spinresults}.

\erwin{Even though we find a single Gaussian feature, it is possible that it consists of several blended CNM components that are impossible to disentangle at this velocity resolution, so that the fitted profile represents their combined absorption.}
Here we describe the
 CNM profile using Eq.~\ref{eq:tau}, with $\tau(v)$ given by
Eq.~\ref{eq:ncomp} for $N=1$.

In fitting the observed emission spectrum, we follow the procedure described in Sect.~3, where for the WNM we use Eq.~\ref{eq:wnm}, with the number of WNM components $K$ part of the fitting process.
With $N=1$, the number of trials is limited to $3^K$.

Determining the number of Gaussian components $K$ is not straightforward. In principle, one could use a goodness-of-fit or information criterion to select the optimal number of components. However, given the limited physical resolution, the observed spectrum is likely only an approximation of the intrinsic emission at the absorber position. It is therefore unclear whether an ``optimal'' $K$ provides a physically meaningful description. As noted by \citet{murray2018} and \citet{chen2025}, even small positional offsets can lead to emission components that do not correspond to those seen in absorption.

\begin{figure}[t]
\centering
\includegraphics[width=0.9\columnwidth]{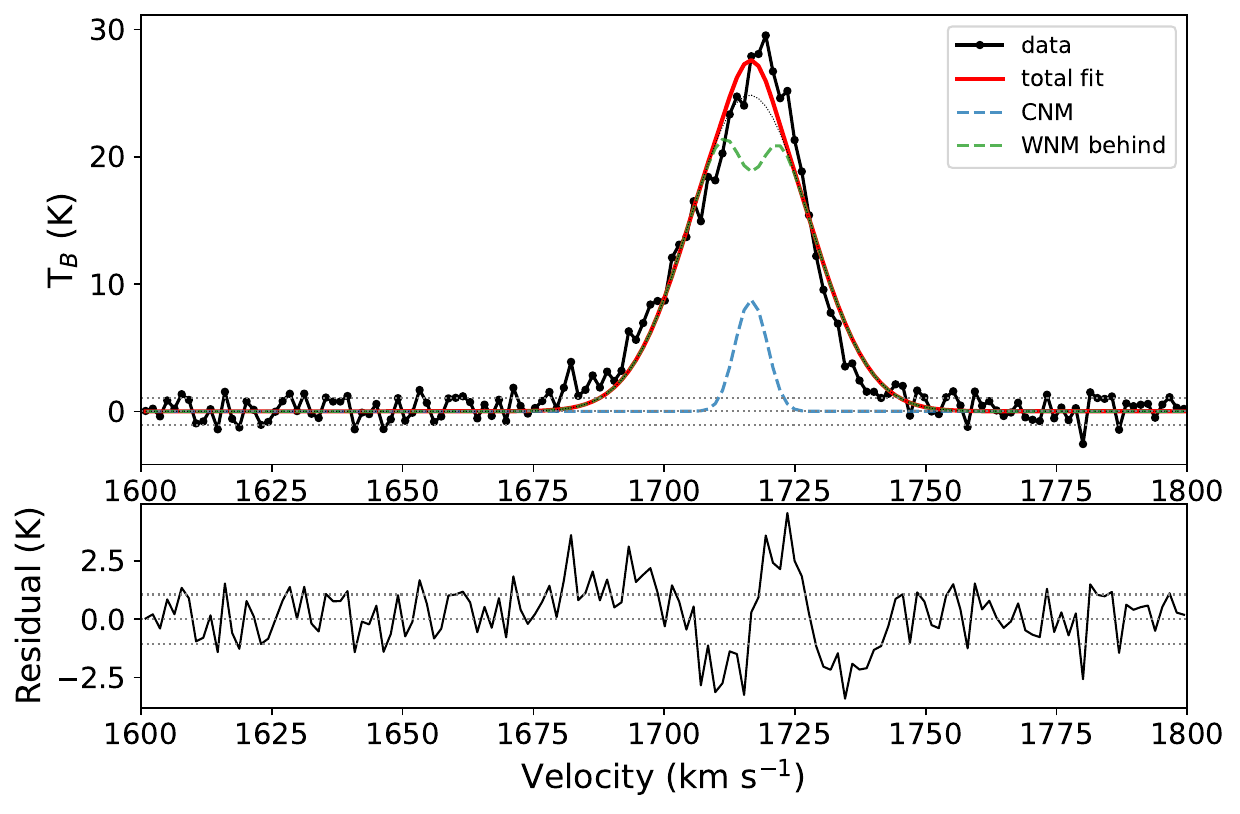}\\
\includegraphics[width=0.9\columnwidth]{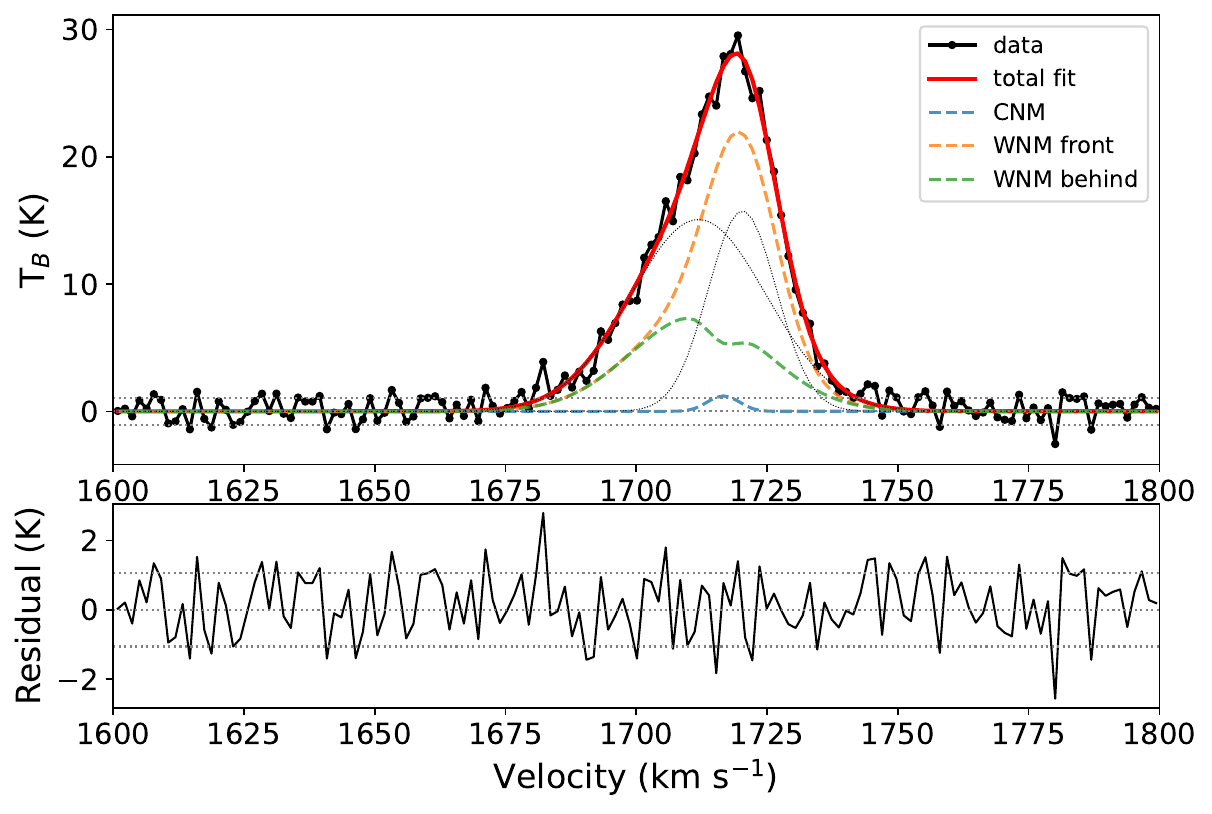}\\
\caption{Average emission spectra derived using the regions shown in Fig.~\ref{fig:N289emissionmom0}. The top panel shows the $K=1$, $F_k=0$ trial. The black line and points indicate the average spectrum. The dashed blue curve shows the CNM emission. The thin dotted black curve shows the intrinsic WNM emission without optical-depth effects. The green curve show the WNM emission, including optical-depth effects, corresponding to gas in behind the absorber. The red curve shows the total (CNM+WNM) emission; residuals are shown in the sub-panel below. Grey lines indicate the zero and $\pm 1\sigma$ levels in both panels.
The bottom panel shows the $K=2$ fit with $F_k = (1.0, 0.5)$. Lines and symbols are as in the top panel. In addition, the orange curve shows the WNM emission in front of the absorber.}
\label{fig:n289spinfits}
\end{figure}

We use the observed spectrum from the two regions discussed above and shown in 
Fig.~\ref{fig:N289emissionmom0}. We explore $K=1$ and $K=2$. 
The best trials are shown in Fig.~\ref{fig:n289spinfits} for both $K$ values. For $K=1$ this is the fit with $F_k=0.0$, for $K=2$ the  fit has $F_k = (1.0, 0.5)$. 
Although only one representative trial is shown for each $K$, the parameters listed in Table~\ref{tab:spinresults} are weighted averages over all $F_k$ combinations (trials) for that $N$ and $K$ (cf.~Sect.~\ref{sec:absem}), with weights inversely proportional to the variance of each trial.

While the $K=2$ fits are formally preferred based on goodness-of-fit, they introduce emission 
components that are not well aligned in velocity with the absorption component and yield unrealistically low spin temperatures. This is a consequence of the mismatch between the spatial scales probed by emission and absorption.  
This effect has been observed even in the MW and the MC, where additional emission components can lead to artificially low inferred spin temperatures.
The uncertainty is therefore dominated by mismatches between emission and absorption, rather than by statistical fitting errors.
For this reason we also consider the $K=1$ fit.
It shows larger residuals, but the central velocity of the WNM component agrees to within 0.3~\kms with that of the cold component. Given this agreement, as well as the low $T_s$ values and mismatched velocities obtained for $K=2$, we conclude that the $K=1$ fit provides the most physically meaningful description of the CNM and WNM at the absorber position.

\begin{table*}
\small
\centering
\caption{Derived properties of the CNM and WNM components along the detected absorption sightlines.}
\begin{tabular}{l @{\hspace{-9pt}}r r r c r r r r r r r }
\hline
\hline
\noalign{\vskip 2pt}
Galaxy/  &  $S_{\rm peak}$ & $\sigma_\tau$ & $f_{\rm CNM}$ & $\langle T_s \rangle$ & Phase &   $T_s$ &  $T_{\rm k,max}$ &  $\tau_{\rm peak}$ or & $v_{\rm peak}$ & $\Delta v_{\rm FWHM}$ & $N_{\rm HI}$ \\
Source  & & & & & & & &$T_{B,{\rm peak}}$ & & & \\ 
& (mJy bm$^{-1}$ ) & & & (K) & & (K) & (K) & & (\kms) & (\kms) & ($10^{20}$ cm$^{-2}$)\\
(1) &  (2) &  (3) &  (4) & (5) & (6) & (7) & (8) &  (9) & (10) &  (11) & (12) \\ 
\hline
\noalign{\vskip 2pt}
NGC 289  & \multicolumn{11}{c}{$K=1$}\\
\hline
\noalign{\vskip 2pt}
 NVSS & 18.3 &  0.022 & 0.093 &	$423 \pm 10$ & CNM   & $24.9 \pm 2.7$ & 1023 & 0.28 & 1716.7 & 6.83 &	1.33 \\
J005245--311503 		 &  & & & & WNM   &  &15763	 & 24.9 & 1716.4 & 26.85 &	12.9 \\

\hline
\noalign{\vskip 2pt}
NGC 289  & \multicolumn{11}{c}{$K=2$}\\
\hline
\noalign{\vskip 2pt}
 NVSS & 	&  & 0.015 &	$423 \pm 10$ & CNM   & $13.9 \pm 1.8$ & 1023 & 0.28 & 1716.7 & 6.84 &	1.33 \\
 J005245--311503 &	  & & & & WNM   &  &5300	 & 15.8 & 1720.3 & 15.57 &	 4.77 \\
  &  & & & & WNM   &  &20314	 & 14.9 & 1711.8 & 30.48 &	8.84  \\
\hline
\hline
NGC 7424& \multicolumn{11}{c}{$K=1$}\\
\hline
\noalign{\vskip 2pt}
 SUMSS          & 45.9 &  0.014 & 0.029 &	$415 \pm 24$ & CNM   & $11.9 \pm 2.7$ & 1445    & 0.077 & 860.6 & 8.13 &	0.03 \\
 J225729–410241 &      &  &       &	             & CNM   & $12.0 \pm 2.7$ &  967   & 0.074 & 880.2 & 6.65 &	0.23 \\
 	              &      &  & &                     & WNM   &                &  26785   & 12.75  & 874.0 & 35.0 &   8.68 \\
\noalign{\vskip 2pt}
\hline
NGC 7424 & \multicolumn{11}{c}{$K=1$}\\
\hline
\noalign{\vskip 2pt}
 SN2001ig & 10.7 & 0.060 & 0.048 &	$847 \pm 31$ & CNM   & $27.4 \pm 2.0$ & 176 & 0.22 & 867.6 & 2.84 &	0.45 \\
 		 &  & & & & WNM   &  &8189 & 20.5 & 865.4 & 21.8 &	8.68 \\
\hline
\end{tabular}
\tablefoot{(1) Name of the foreground galaxy and the background source. (2) Peak flux of the continuum source. (3) Optical-depth rms uncertainty in absorption spectra. (4) Fraction of CNM mass compared to total \HI mass. (5) Density-weighted mean spin temperature. (6) Phase of the \HI. (7) Spin temperature (for CNM only). (8) Maximum kinetic temperature. (9) Peak optical depth (for CNM) or peak brightness temperature (for WNM). (10) Velocity of the peak optical depth or brightness temperature. (11) FWHM velocity width. (12) \hi column density of this component. }
\label{tab:spinresults}
\end{table*}

We now derive the properties of the CNM and WNM along this sight line.
From Eqs.~\ref{eq:tau} and \ref{eq:nhicnm}, we find the column density corrected for optical depth effects:
\begin{equation}
N_{\rm \ion{H}{i},thick} = 1.823 \times 10^{18} \int \frac{T_B \tau}{1-e^{-\tau}} dv.
\end{equation} 
This can be compared with the optically thin expression (Eq.~\ref{eq:nhiwnm}). These values are listed in Table~\ref{tab:spinresults}.
For this line of sight the ratio of these column densities is 1.036,
\erwin{though this number is likely to be a lower limit, given the resolution effects. We will discuss further in Sect.~\ref{sec:detections}.}


\begin{table}
\small
\centering
\caption{\HI column densities along the absorption lines with and without optical-depth corrections}
\begin{tabular}{l r r r}
\hline
\hline
\noalign{\vskip 2pt}
Source  &  $N_{\rm HI,thick}$ & $N_{\rm HI,thin}$ & thick/thin \\

 & ($10^{20}$ cm$^{-2}$) & ($10^{20}$ cm$^{-2}$) & \\
(1) &  (2) &  (3) &  (4) \\ 
\hline
\noalign{\vskip 2pt}
NVSS J005245--311503 & 14.22 & 13.73 & 1.036 \\
 SUMSS   J225729--410241        & 8.82 & 8.71 & 1.012 \\
  SN2001ig      & 9.60 & 9.46 & 1.014 \\
\hline
\end{tabular}
\tablefoot{(1) Name of background source. (2) \hi column density after correction for optical depth. (3) \hi column density assuming optically thin gas. (4) Ratio of \HI column densities from columns (2) and (3).}
\label{tab:nhithick}
\end{table}

\subsection{Absorption in NGC 7424}

\subsubsection{Properties of the galaxy and the continuum source}

We detect \HI absorption in two sight lines in NGC 7424 (J2257--41). The galaxy is  late-type and barred, with an \hi disk measuring ${\sim}18$ kpc. It hosts the supernova SN2001ig \citep{ryder2004,ryder2006,ryder2018} and contains a number of strong ultra-luminous X-ray sources (e.g., \citealt{soria2006}).

As in NGC~289, the \HI and stellar masses ($4.0 \times 10^9$ and $3.6 \times 10^9\,\msun$, respectively) are nearly equal \citep{deblok2024}, indicating a gas-rich system.
Though located in the vicinity of the IC 1459 group \citep{serra2015}, it is considered to be a field galaxy \citep{yahil1977,kourkchi2017}. 
We assume a distance of 7.9 Mpc \citep{leroy2019}, corresponding to $1'' = 0.04$ kpc.
The first detailed \HI observations (obtained with ATCA) were presented in \citet{reeves2015}.  NGC 7424 was also observed in \HI with KAT-7 by \citet{sorgho2019}. They derived its rotation curve, showing a rapid rise with an extended flat part at around $\sim 140$ \kms. 

\begin{figure}[ht]
\centering
\includegraphics[width=0.9\columnwidth]{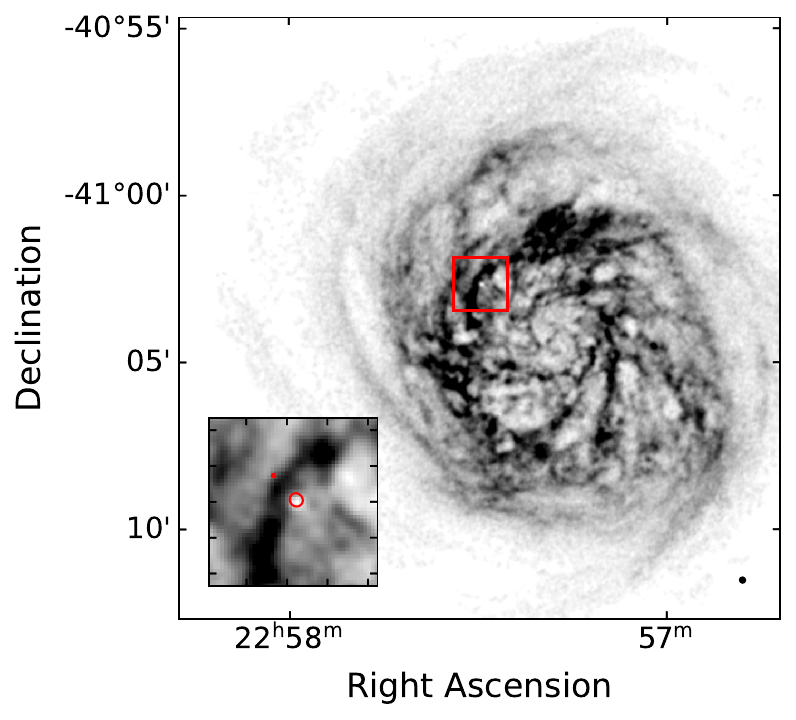}
\caption{The \rzero\ zeroth-moment map of NGC~7424. The red box ($1.6' \times 1.6'$) marks the position of the absorption features. The bottom-left inset shows a zoom of this region. The two red ellipses indicate the continuum sources: the larger ellipse marks SUMSS~J225729$-$410241 and the smaller ellipse SN2001ig. The \rzero\ beam is shown in the lower right.}
\label{fig:2257mom0}
\end{figure}

The absorption spectra are shown in the central and bottom panels of Fig.~\ref{fig:detectionspectra}. 
The detection toward SUMSS J225729--410241 shows two absorption features at different velocities, likely associated with distinct CNM complexes, and is visible in the \rzero zeroth-moment map (Fig.~\ref{fig:2257mom0}).
 \citet{reeves2015} also searched for \hi absorption against SUMSS J225729–410241, but did not detect any. 
The absorption feature is visible also in the {\tt r05\_t00} moment map, but not in the {\tt r10\_t00} or lower-resolution maps.

A second absorption spectrum is detected toward SN2001ig, where absorption occurs against a source within the galaxy rather than a background source.
It is not clear whether the absorption is associated with the supernova environment, or whether it is intervening absorption from the \HI disk. The SN2001ig absorption feature is not  visible in the zeroth-moment map due to the presence of \hi emission at higher and lower velocities.

Properties of the two continuum sources are listed in Table~\ref{tab:contcat}.
For SUMSS J225729–410241, \citet{reeves2015} find a peak flux of $44.4 \pm 1.7$ \mJybeam at 1.4 GHz (based on their highest resolution data). This is consistent with our measured peak flux of $45.9$\,\mJybeam. The source is unresolved and known to be a background QSO \citep{ryder2004,soria2006}. 
Catalogued as WISEA J225729.67--410239.4, it is also visible in the WISE W1 and W2 bands, but not in W3 and W4. 
The Legacy Survey shows a counterpart that is significantly redder than the surrounding stellar and star-forming population of NGC~7424. 

The second sight line with absorption is toward SN2001ig. This supernova is classified as a Type IIb, and was notable for the flux modulations in the early aftermath of the explosion, possibly indicating the presence of circumstellar material \citep{ryder2004}.

Our measured flux density of 10.8\,mJy is nearly twice the value of $5.1 \pm 0.5$\,mJy measured $\sim700$ days after the explosion \citep{ryder2004}.
SN2001ig was observed serendipitously by \citet{reeves2015}. They record 1.4 GHz fluxes of
$9.0 \pm 0.6$ and $8.8 \pm 0.7$ mJy, for observations done in October 2011 and June 2013 respectively, also suggesting a brightening compared to a decade earlier. 
If physical, this increase could be due to shock-powered emission associated with the transition from the supernova to the remnant phase.
SN2001ig is unresolved in our data.
Both sources have  been detected in X-ray using \emph{Chandra} observations \citep{soria2006}.
The projected galactocentric distance of the sources is $\sim 6.0$ kpc.

\subsubsection{The absorption spectra\label{sec:n7424absorption}}

We apply the procedure described in Sect.~\ref{sec:absorption} to construct an \HI-emission-free cube.
We used the same robustness parameter of $-1$, but impose a larger baseline flagging limit of 1500\,m, due to the more prominent \HI emission in the cube. A complicating factor is the presence of two bright continuum sources (with peak fluxes of $\sim 0.3$ Jy beam$^{-1}$) in the outer primary beam, about $45'$ from the center. These cause low-level ripples that become prominent in the \HI-emission-free cube (where the continuum has not been subtracted), requiring adjustments to the fitting to mitigate their impact.
The final emission-free cube has a noise level of 0.64 \mJybeam.

\begin{figure}[th]
\centering
\includegraphics[width=0.9\columnwidth]{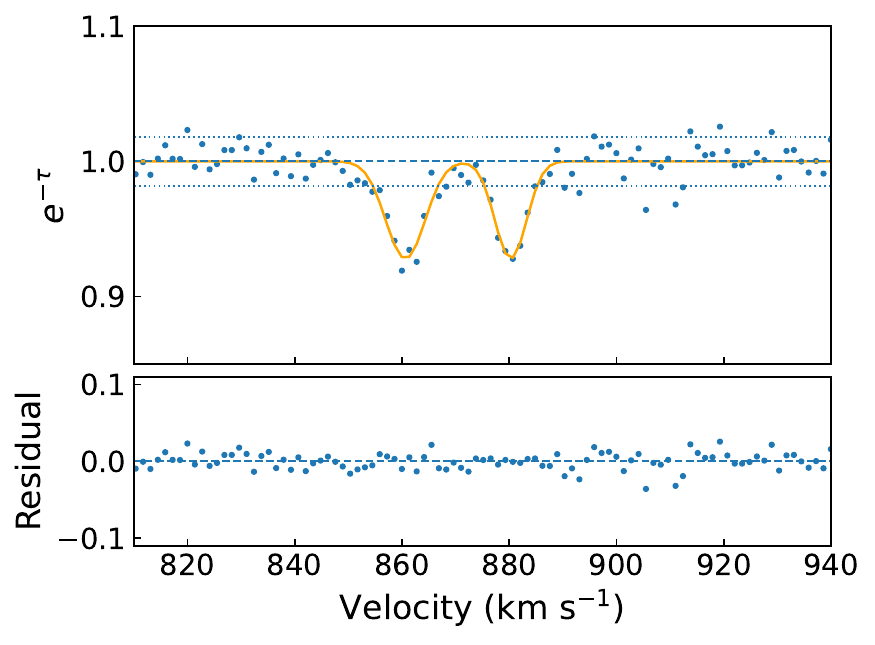}
\includegraphics[width=0.9\columnwidth]{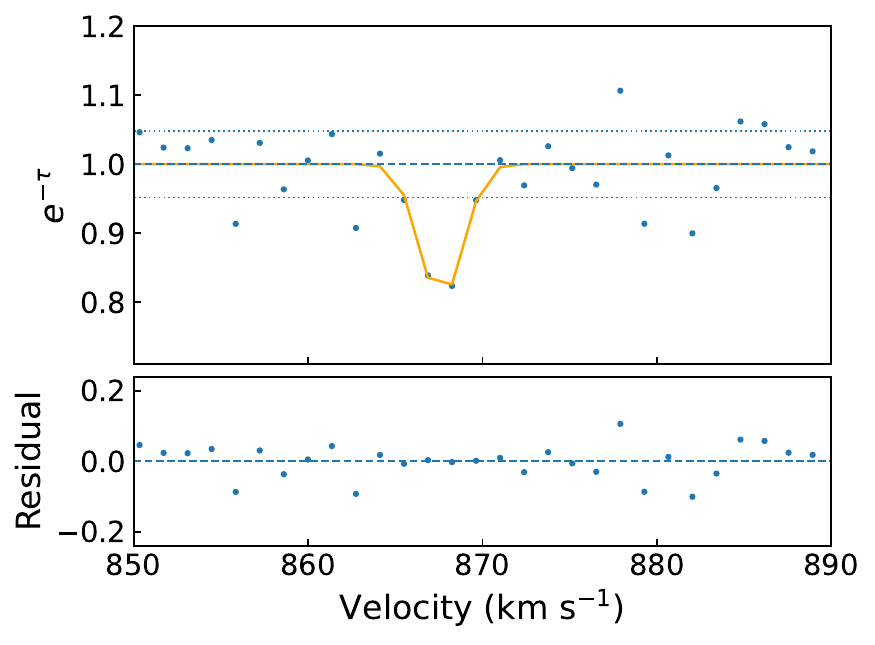}
\caption{Top: intrinsic absorption spectrum in NGC~7424 towards SUMSS~J225729$-$410241 derived from the absorption-only data cube. The upper sub-panel shows the spectrum as optical depth $\tau$. The dashed line marks zero optical depth and the dotted lines indicate the $\pm 1\sigma$ levels. Two Gaussian components are overplotted. The lower sub-panel shows the residuals with respect to the fit. Bottom: absorption spectrum towards SN2001ig. Lines and symbols are as in the top panel.}

\label{fig:J2257_SN_absorption}
\end{figure}

The absorption toward SUMSS J225729--410241 shows two components
(Fig.~\ref{fig:J2257_SN_absorption}). In quantifying the component parameters, the effect of the ripples meant including a quadratic baseline in the fit, and limiting the fitting range to 820--920 \kms. The two components have similar properties. The peak optical depths are $\tau_0 = (0.076 \pm 0.007, 0.074 \pm 0.007)$. The central velocities $v_0$ are (860.6, 880.3)\,\kms. The velocity dispersions of the components are $\sigma_0 =  (3.46, 2.83)$\,\kms [FWHM = (8.13, 6.65)\, km s$^{-1}$].
Central velocities and velocity dispersions have very small formal fitting uncertainties. 
The low-velocity absorption component shows tentative evidence for a small wing on its low-velocity side, but given the presence of the ripple it is difficult to assign a statistical significance to it. 
  
The absorption spectrum towards SN2001ig is much fainter (Fig.~\ref{fig:J2257_SN_absorption}). The continuum ripples necessitate the inclusion of a linear baseline fit and a narrow fitting range of 850 to 890 \kms. The absorption spectrum is shown in Fig.~\ref{fig:J2257_SN_absorption} and can be described by a single Gaussian component. The most notable property is its narrow velocity width: we find a velocity dispersion of $1.21 \pm 0.36$ \kms (FWHM $= 2.84 \pm 0.85$ \kms), close to the velocity resolution of our data (1.4 \kms). The peak optical depth is high, with a value of $\tau_0 = 0.22 \pm 0.05$. The central velocity is $v_0 = 867.6 \pm 0.3$ \kms.
These parameters are summarized in Table~\ref{tab:spinresults}.
  
\subsubsection{Emission spectra}

\begin{figure}[th]
\centering
\includegraphics[width=0.9\columnwidth]{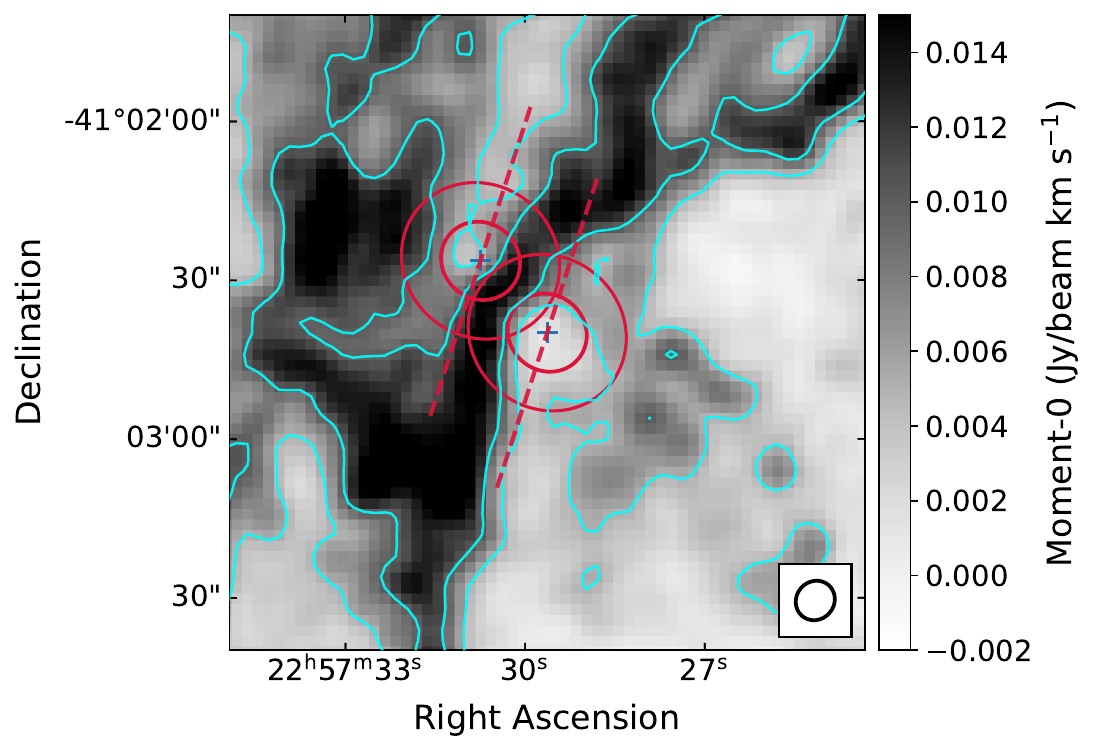}
\caption{Zeroth-moment map of the region around the NGC~7424 absorbers in grayscale. The map was created by collapsing channels of the \rzero\ data cube between 866.4 and 867.6 \kms, corresponding to the velocity range of the SN2001ig absorption feature. No masking was applied. Cyan contours show the \hi\ column density at 0.005 and 0.01 Jy beam$^{-1}$ \kms\ (corresponding to $1.1 \times 10^{20}$ and $2.2 \times 10^{20}$ \cm, respectively). Crosses indicate the positions of the continuum sources. Annuli with inner radius one beam width and outer radius two beam widths are overplotted and bisected by dashed lines at a position angle of $-18^\circ$. The \rzero\ beam is shown in the lower right.}

\label{fig:N7424emissionmom0}
\end{figure}

We again take a detailed look at the  spatial variation of the \HI column density distribution near the absorbers as well as the velocity field to gauge any possible impact on the derivation of averaged profiles.
Figure~\ref{fig:N7424emissionmom0} shows the zeroth-moment map of the area around the absorbers, created by summing the three channels between 866.4 and 867.6 \kms. This narrow velocity range was chosen to maximise the visibility of the SN2001ig absorption feature. The SUMSS J225729–410241 absorption is also visible despite the velocity range covering only a small part of the feature (cf.\ the absorption spectra in Fig.~\ref{fig:J2257_SN_absorption}). 

\begin{figure}[t]
\centering
\includegraphics[width=0.9\columnwidth]{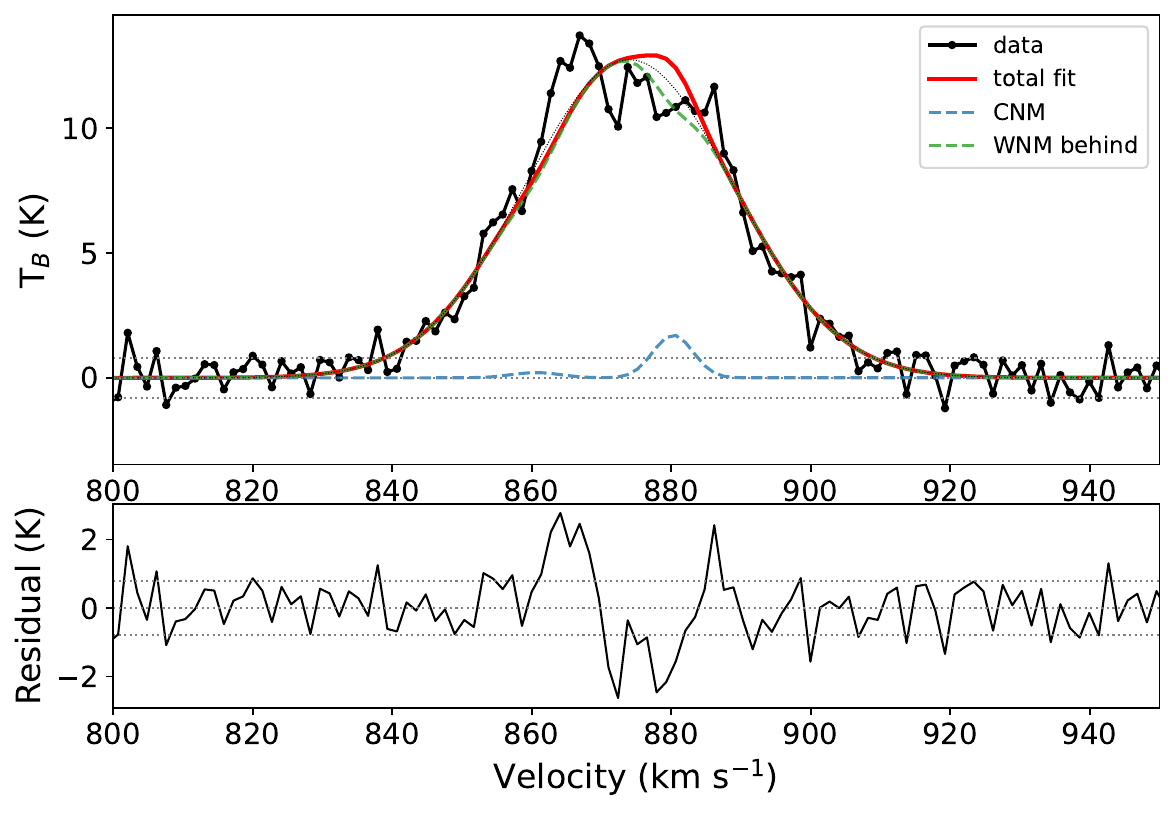}\\
\includegraphics[width=0.9\columnwidth]{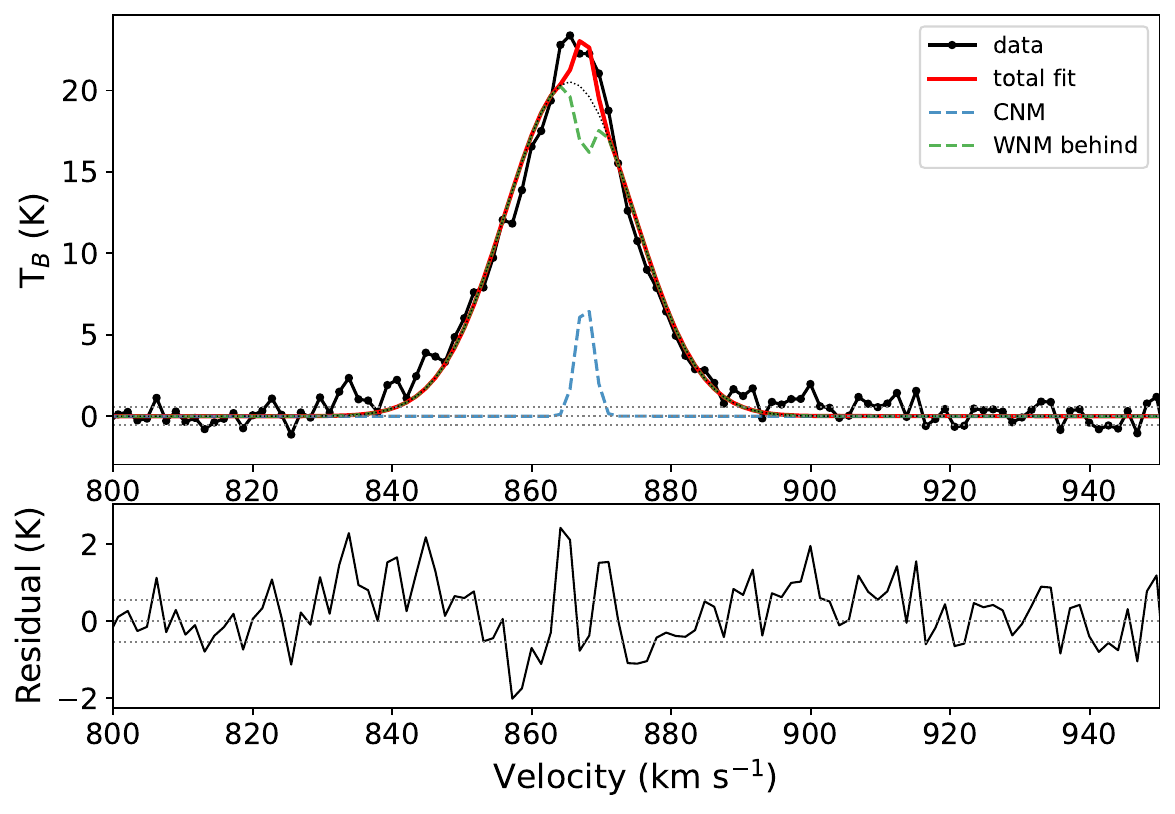}\\\caption{Average emission spectra derived using the regions shown in Fig.~\ref{fig:N7424emissionmom0}. The top panel shows the spectrum towards SUMSS~J225729$-$410241, derived from the western half of the annulus. The $K=1$, $F_k=0$ trial is shown. The black line and points indicate the average spectrum. The dashed blue curves show the two CNM components. The thin dotted black curve shows the intrinsic WNM emission without optical-depth effects. The green curves show the WNM emission including optical-depth effects, corresponding to gas behind the absorber. In the trials displayed here there is no WNM component in front of the absorbers. The red curve shows the total (CNM+WNM) emission; residuals are shown in the panel below. Gray lines indicate the zero and $\pm 1\sigma$ levels in both panels.
The bottom panel shows the same for SN2001ig, using the eastern half of the annulus. The $K=1$, $F_k=0$ trial is shown; lines and symbols are as in the top panel.}
\label{fig:N7424spinfits}
\end{figure}

The velocity gradient across the absorbers is only a few \kms, i.e., much smaller than for NGC 289.
However, they are located on opposite steep edges of a high–column density \HI spiral arm. 
As was the case for NGC~289, averaging over a full annulus (which includes the spiral arm) yields non-representative emission spectra.
To mitigate this we define an annulus around each source, with an inner radius equal to one beam size and an outer radius of two beams.
We divide each annulus into two halves using lines at a position angle of $-18\degree$, approximating the orientation of the spiral arm.
For SUMSS J225729–410241, we use the western (right) part of the annulus to determine the emission spectrum. For SN2001ig we use the eastern (left) part. This avoids including the high-column-density spiral arm in the averaged spectra, \erwin{as the absorbers are located away from these high column densities.}
The resulting averaged emission spectra are shown in Fig.~\ref{fig:N7424spinfits}.

\subsubsection{Analysis of the spectra}

We fit the absorption and emission spectra using the same procedure as
described in Sect.~\ref{sec:analysis}. For SUMSS J225729--410241 we
have two CNM components, i.e., $N=2$.  For the WNM along this line of
sight a single component ($K=1$) is preferred. The combined spectrum
is shown in Fig.~\ref{fig:N7424spinfits} where we show the $F_k = 0$
trial.  Models with $K=2$ or $K=3$ do not improve the fit.  The figure
also illustrates the difficulty of determining the intrinsic emission
spectrum in a complex environment with limited spatial
resolution. With this single WNM component, the low-velocity CNM
component yields a low $T_s$ value. \erwin{Although the detailed emission
profiles and derived CNM and WNM properties vary somewhat with the choice
of emission regions, all choices consistently yield a low $T_s$,
reinforcing the earlier conclusion that matching emission and
absorption spectra is intrinsically difficult given the different
spatial scales probed.}

For SN2001ig we have a single CNM component $(N=1)$ and find that the emission spectrum is best fit with a single WNM component ($K=1$). The best-fitting trial with $F_k=0$ is shown in Fig.~\ref{fig:N7424spinfits}. The WNM has a velocity dispersion of ${\sim}9$ \kms which is a value typically found in the outer parts of \HI disks. Models with $K=2$ or $K=3$ do not lead to quantitatively better fits. The corrections for optical depth are small (but again likely lower limits). 
All parameters are summarized in Table~\ref{tab:spinresults}.

\section{Discussion}

\subsection{Absorption limits\label{sec:abslimits}}

\begin{figure}[t]
\centering
\includegraphics[width=0.9\columnwidth]{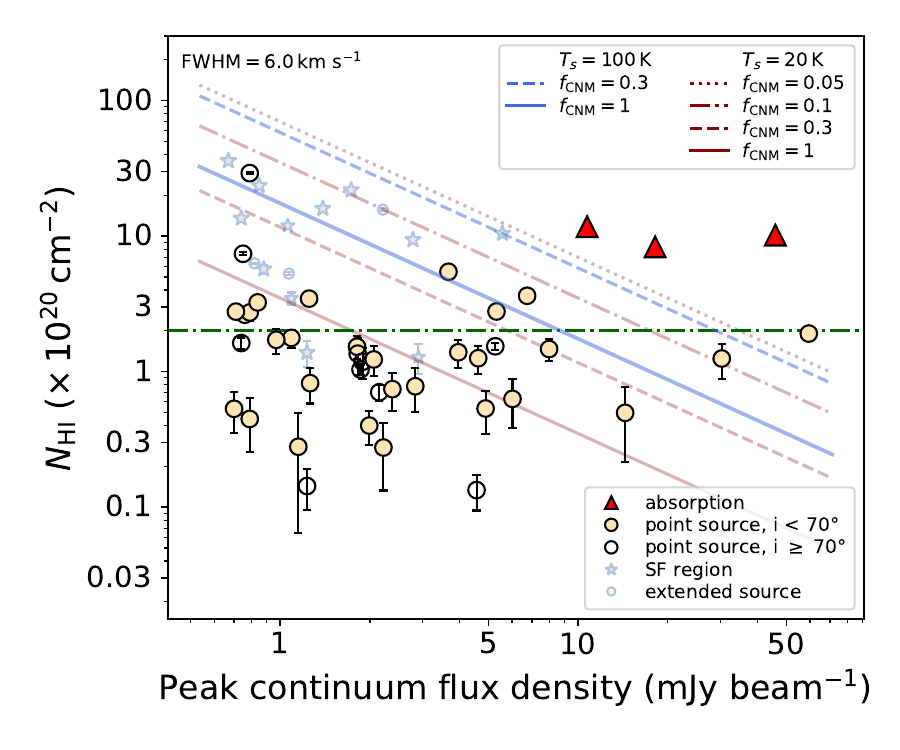}\\
\includegraphics[width=0.9\columnwidth]{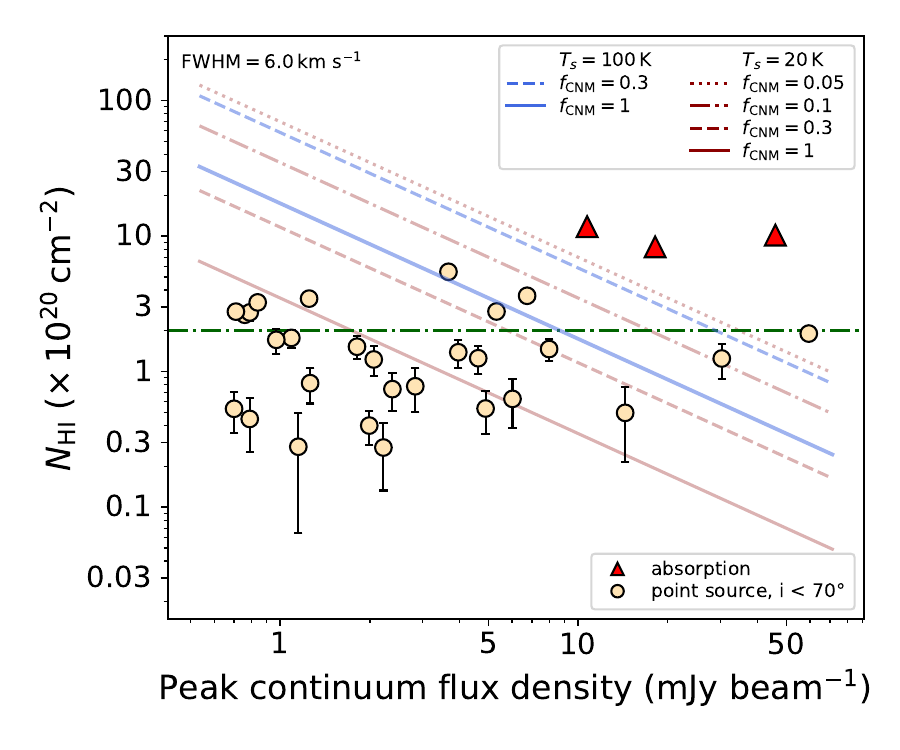}\\
\caption{Top: inclination-corrected \hi\ foreground column density plotted against the peak flux density of the background continuum sources. Red triangles indicate absorption detections. Circles indicate non-detections against unresolved, non-H$\alpha$ sources; filled circles correspond to $i<70\degree$, open circles to $i\geq 70\degree$. Small stars indicate H$\alpha$ star-forming regions, and small circles resolved sources not associated with H$\alpha$. The filled circles define the ``clean'' sample of unresolved, non-H$\alpha$, low-inclination sources. The horizontal line marks the \citet{kanekar2011} threshold at $2 \times 10^{20}$ \cm. The dashed diagonal lines show the $3\sigma$ absorption detection limits for different $T_s$ and $f_{\rm CNM}$ values. Bottom: same as top, but showing only the clean sample for clarity.}
\label{fig:peaknhiresult}
\end{figure}

Figure~\ref{fig:peakNHI} shows that absorption detections are confined to one part of the diagram, approximately defined by $S_{\rm peak} \gtrsim 9$ \mJybeam and $N_{\ion{H}{i}} \gtrsim 7 \times 10^{20}$ \cm. To relate the observed \HI column densities to the intrinsic gas properties, we correct them for galaxy inclination. The resulting distribution is shown in the top panel of Fig.~\ref{fig:peaknhiresult}, where we again indicate extended sources, sources associated with H$\alpha$, and sources located behind high-inclination galaxies.
The bottom panel of Fig.~\ref{fig:peaknhiresult} shows only the clean sample, which we consider below.

One factor that can determine whether absorption is detectable is the presence of a CNM phase. In studies of Galactic absorption spectra, \citet{kanekar2011} find a strong decrease in $T_s$ at $N_{\rm \ion{H}{i}} \simeq 2 \times 10^{20}$ \cm, with higher column densities showing significantly lower temperatures. This transition is interpreted as the onset of CNM formation. Below this threshold, strong absorption is not generally expected.

Figure~\ref{fig:peaknhiresult} shows that about two-thirds of the clean sources have foreground column densities below $2 \times 10^{20}$ \cm, consistent with the absence of absorption. However, several sources with higher column densities also show no absorption. Reconciling these with a column-density threshold alone would require increasing the threshold to $\sim 7 \times 10^{20}$ \cm, well within the regime where CNM is expected to be present (cf.\ Fig.~1 in \citealt{kanekar2011}). This suggests that a column-density threshold alone cannot account for the lack of absorption.

This remaining discrepancy is likely related to the optical-depth sensitivity of the data.
To derive limits on the \hi column density from absorption non-detections, we consider the optical-depth sensitivity. The optical depth is given by
$\tau(v) = -\ln({S(v)}/{S_{\rm peak}})$,
where $S(v)$ is the observed flux density toward the background continuum source in a single spectral channel at velocity $v$, and $S_{\rm peak}$ is the continuum level measured from line-free channels.

The uncertainty in the optical depth per channel follows from standard error propagation, assuming that the uncertainty in the continuum level is negligible. This gives
$\sigma_\tau = \sigma_I/{S(v)}$,
where $\sigma_I$ is the rms noise per spectral channel in $S(v)$. In the optically thin limit, this expression plays the same role for the uncertainty as Eq.~\ref{eq:deltaS} does for the optical depth itself.

In the optically thin limit, $S(v) \simeq S_{\rm peak}$ and the uncertainty reduces to
\begin{equation}
\sigma_\tau \simeq \frac{\sigma_I}{S_{\rm peak}}.
\end{equation}
To convert the corresponding  peak optical-depth limit $\tau_{\rm peak}^{\rm lim}=n\sigma_\tau$ into an H\,\textsc{i} column density limit, we assume a Gaussian absorption profile with $\mathrm{FWHM} = 2.35\,\sigma$. The velocity-integrated optical depth is then
$
\int \tau(v)\,dv = 1.064\,\tau_{\rm peak}\,\mathrm{FWHM}.
$
Using Eq.~\ref{eq:nhicnm} with $c_f = 1$, we obtain
\begin{equation}
N_{\rm \ion{H}{i}}^{\rm lim}(S_{\rm peak})
=
1.94 \times 10^{18}\,
T_s\,
n\,
\mathrm{FWHM}\,
\frac{\sigma_I}{S_{\rm peak}}.
\label{eq:hilimit}
\end{equation}
In Sect.~\ref{sec:absorption} and \ref{sec:n7424absorption}, we measured $\sigma_I = 0.41$ \mJybeam for NGC~289 and $0.64$ \mJybeam for NGC~7424. We adopt a representative value of $\sigma_I = 0.5$ \mJybeam and assume $n=3$. We further adopt $\mathrm{FWHM} = 6$~\kms, corresponding to the average width of the CNM components.

We explore two values for the spin temperature. The first is $T_s = 20$\,K, corresponding to the average of our CNM measurements. This lies at the lower end of values derived in the Magellanic Clouds (cf.\ \citealt{dempsey2022,chen2025}), and, as discussed in Sect.~\ref{sec:detections}, may underestimate the true value. We also consider $T_s = 100$\,K, commonly adopted in the literature.

For both $T_s$ values, we calculate the $3\sigma$ detection limits as a function of continuum peak flux. As we compare these limits to the total observed \HI column densities, we also include a range of $f_{\rm CNM}$ values. The resulting limits are shown in Fig.~\ref{fig:peaknhiresult}.

For both spin-temperature assumptions, the non-detections are consistent with a lack of optical-depth sensitivity. For $T_s = 20$\,K, this requires $f_{\rm CNM} \sim 0.1$, comparable to values found in the SMC \citep{dempsey2022}. For $T_s = 100$\,K, $f_{\rm CNM} \sim 0.3$, typical of the Milky Way \citep{mccluregriffiths2023}, is sufficient. Reasonable combinations of $T_s$ and $f_{\rm CNM}$ therefore explain most non-detections in terms of limited optical-depth sensitivity.

One exception is the spectrum towards the brightest source, SUMSS~J030941$-$410006 behind ESO~300-G014 (J0309$-$41). Although it lies above the nominal optical-depth sensitivity limits, its column density is close to the $2 \times 10^{20}$ \cm threshold \citep{kanekar2011}. Inspection of the spectrum reveals no convincing absorption features (Fig.~\ref{fig:nodetections}). A narrow negative spike near 960 \kms is inconsistent with a real feature, as it disappears after Hanning smoothing and lies outside the velocity range of the \HI emission at that location ($911 \pm 15$ \kms, derived from the first- and second-moment maps).
We therefore conclude that the absence of absorption towards most continuum sources is primarily driven by limited optical-depth sensitivity.

\citet{reeves2015,reeves2016} present a search for intervening absorption in 16 nearby gas-rich galaxies using ATCA, including observations of SUMSS~J225729$-$410241 behind NGC~7424. Despite additional observations, no absorption was detected by them. At first glance this may appear surprising, as this is the most prominent absorption feature in the MHONGOOSE sample.
\citet{reeves2015} report a $3\sigma$ upper limit on $\tau_0$ of 0.09 for their highest-resolution and deepest NGC~7424 data (ATCA-6C array)\footnote{The values of $\tau_{\rm peak}$ and $\int \tau\,dv$ in Table~8 of \citet{reeves2015} are consistent only if $\tau_{\rm peak}$ is interpreted as the listed value, rather than as a percentage as implied by the column header.}. Assuming a FWHM of 10 \kms, they derive a $3\sigma$ upper limit of $\int \tau\, dv = 0.99$ \kms. Using a width of 6 \kms instead, comparable to our CNM components, gives an integrated optical-depth limit of 0.59 \kms.
This can be compared with the average integrated optical depth we measure for NGC~7424, which is also $0.59$ \kms. Our detected absorption profiles therefore lie close to the sensitivity limit of the \citet{reeves2015} data. Their non-detection is thus consistent with the properties of the absorption detected here.

\subsection{Stacking}

To search for additional absorption in the two galaxies with detected absorption features, we constructed stacked spectra from the \rzero emission cubes at the positions of background continuum sources identified with \texttt{PyBDSF} down to $5\sigma_{\rm cont}$, where $\sigma_{\rm cont}$ is the noise in the primary-beam-corrected continuum maps. This selection includes sources with flux densities below 0.7~mJy. Within the area covered by the \rzero zeroth-moment maps, this yields 120 background sources for J0052$-$31 and 67 for J2257$-$41. The rms noise in the continuum images prior to primary-beam correction is \SI{3.0}{\micro Jy\,beam^{-1}} for both fields.

\begin{figure}[t]
\centering
\includegraphics[width=0.9\columnwidth]{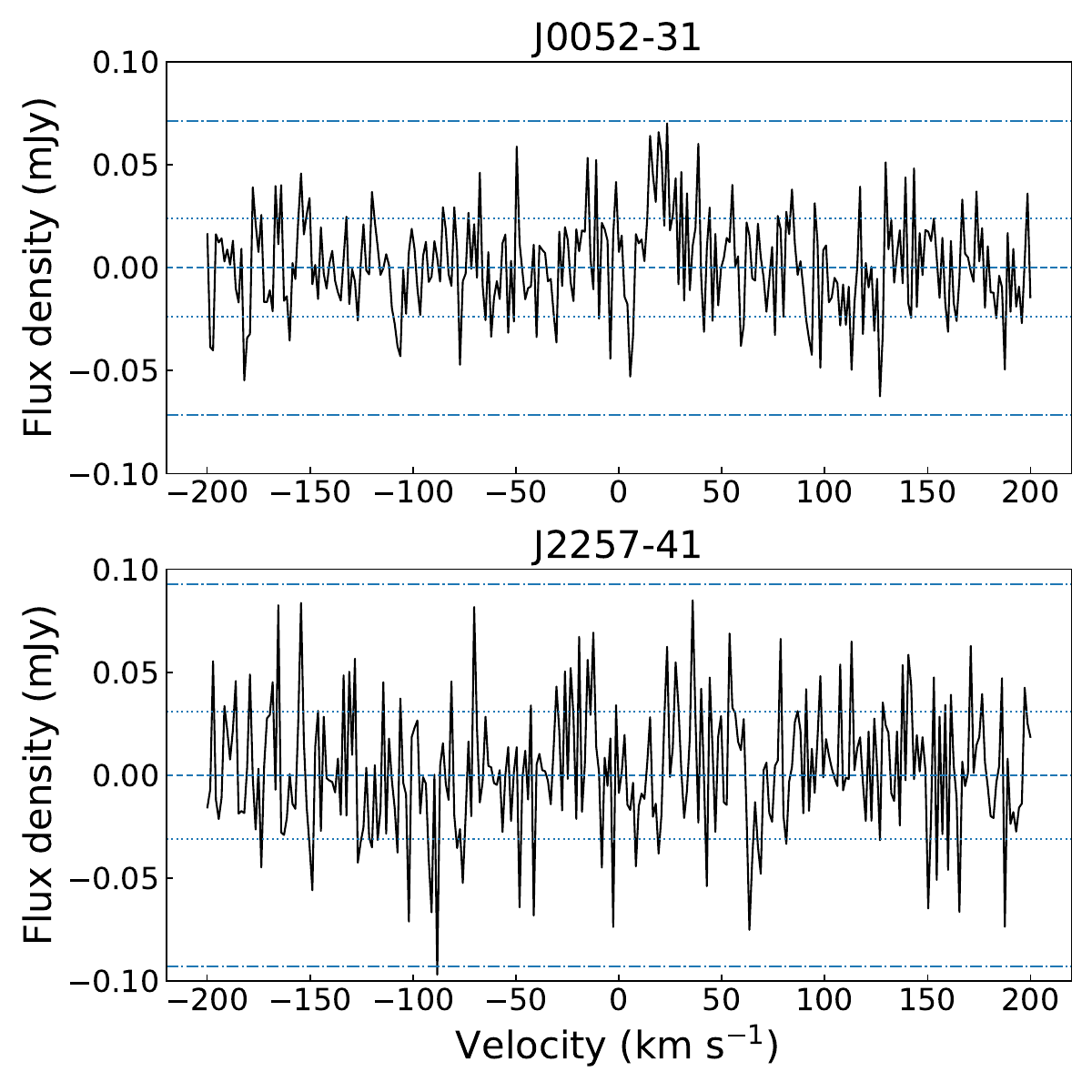}\\
\caption{Top: median stacked spectrum of 120 velocity-aligned sight lines towards continuum sources in NGC~289 (J0052$-$31). The horizontal lines indicate the zero level (dashed), and the $\pm 1\sigma$ (dotted) and $\pm 3\sigma$ (dash-dotted) levels. Bottom: same for NGC~7424 (J2257$-$41), based on 69 sight lines.}

\label{fig:stacked}
\end{figure}

We aligned the spectra to a common velocity frame using the first-moment velocities at each position, assuming that these trace the velocities of the absorption. The \hi\ emission was removed by subtracting Gaussian profile fits with the following constraints: positive amplitude, central velocity within $\pm 30$ \kms, and velocity dispersion between 2.8 \kms\ (two channels) and 50 \kms. The stacked spectra were constructed as the median of the residual spectra from the 120 and 67 directions, respectively. The resulting rms noise levels are $\sim$\SI{20}{\micro Jy} and $\sim$\SI{30}{\micro Jy} for J0052$-$31 and J2257$-$41, respectively. These spectra are shown in Fig.~\ref{fig:stacked} and show no absorption features at the $3\sigma$ level.
The stacked spectrum of J0052$-$31 shows residual emission near $+25$ \kms\ at a level just below $3\sigma$. This is likely due to non-Gaussian line shapes in some of the individual spectra.

For comparison, we constructed stacked spectra at random positions within the \rzero moment maps, i.e., without associated background continuum sources. These spectra show similar noise levels and no evidence for absorption features, consistent with the results above. A more detailed analysis, including (optical-depth-weighted) stacking across the full MHONGOOSE sample, is beyond the scope of this paper.

\subsection{Detections\label{sec:detections}}

\citet{pingel2024} present a detailed study of \HI\ absorption in the disk of the Local Group galaxy NGC~6822. Using high-resolution Local Group L-Band Survey (LGLBS) data \citep{koch2025}, they search for absorption toward 18 sources with peak flux densities $>0.65$ \mJybeam, detecting five CNM components along two lines of sight. No absorption is found toward the remaining sources.
In terms of angular resolution, velocity resolution, and sensitivity, this study is comparable to our work. They report a beam of $7.0'' \times 5.2''$, a channel spacing of 0.4 \kms, and a channel noise of 7~K. This corresponds to 0.42 \mJybeam per 0.4 \kms channel, or 0.22 \mJybeam when smoothed to 1.4 \kms, identical to our noise per 1.4 \kms channel.

Table~\ref{tab:compare} compares the CNM properties. While the ranges in linewidth and spin temperature are similar, both the CNM fraction and CNM column density are systematically lower in our sample. \citet{pingel2024} also find lower CNM fractions than in the MW, MCs, M31, and M33, and interpret this as consistent with evidence that the CNM fraction decreases with metallicity (e.g., \citealt{wolfire1995}).
Applying this interpretation to our results would imply metallicities for NGC~289 and NGC~7424 lower than the $\sim0.2$ solar value of NGC~6822, which is unlikely given their higher masses. \citet{walsh1997} find NGC~289 to be approximately solar metallicity, while NGC~7424 is likely about half solar \citep{walcher2006,modjaz2011}. The systematically lower $f_{\rm CNM}$ and $N_{\rm \ion{H}{i}}^{\rm CNM}$ are therefore more likely due to resolution effects associated with the larger distances of our galaxies (21.5 and 7.9 Mpc versus 0.5 Mpc).

\begin{table}
\small
\centering
\caption{Comparison of CNM properties between this work and \citet{pingel2024}}
\begin{tabular}{l r r }
\hline
\hline
\noalign{\vskip 2pt}
Quantity & this work & \citet{pingel2024} \\
\hline
\noalign{\vskip 2pt}
$T_s$  (K)                  & 11.9 -- 27.4        &  16.3 -- 46.6  \\
$N_{\rm \ion{H}{i}}^{\rm CNM}$  ($\times 10^{20}$\,\cm)        & 0.03 -- 1.33        &  1.2 -- 8.6     \\
$\Delta v_{\rm FWHM}^{\rm CNM}$ \kms & 2.8 -- 8.1          &  2.8 -- 4.4      \\  
$f_{\rm CNM}$                      & 0.093, 0.029, 0.048 &  0.16, 0.37 \\
\hline
\end{tabular}
\label{tab:compare}
\end{table}

The CNM column density is derived from the absorption spectrum (Eq.~\ref{eq:nhicnm}) and represents a pencil-beam measurement that is independent of galaxy distance. In contrast, quantities such as the total \hi\ column density and $f_{\rm CNM}$ rely on emission spectra, and depend on how representative the emission is of the absorption sight line. These spectra are typically extracted over areas comparable to or larger than the synthesized beam. For Local Group galaxies this corresponds to parsec scales (e.g.\ $\sim$15~pc for NGC~6822; \citealt{pingel2024}), whereas for more distant systems the same angular resolution corresponds to kiloparsec scales (0.3 and 0.8~kpc for NGC~7424 and NGC~289, respectively).

In the following we adopt a representative physical scale of 0.6~kpc. If the ISM were spatially uniform, changes in scale would have no impact. However, the ISM is structured, with the CNM occupying only a fraction of the projected area, while the WNM is more smoothly distributed (e.g., \citealt{soler2022,smith2023,mccluregriffiths2023}).
The structure of the \hi\ can be characterized by the projected fractal dimension $D_p$, which describes how the area $A(L)$ covered by \hi\ structures scales with size $L$ as $A(L)\propto L^{D_p}$. A uniform medium has $D_p=2$, while $D_p<2$ indicates a clumpy or filamentary morphology. \citet{stanimirovic1999} measure $D_p\simeq1.5$ for the SMC over scales from tens of parsecs to several kiloparsecs. Although this formally applies to total \hi, small-scale structure is dominated by cold gas, while the WNM provides a smoother background. Similar values are found for molecular clouds in the MW.

For a beam of size $L$, the projected CNM covering fraction can be written as
\begin{equation}
C_{\rm area}^{\rm CNM}(L) = \frac{A_{\rm CNM}(L)}{L^2} \propto L^{D_p-2}.
\end{equation}
With $D_p\simeq1.5$, this gives $C_{\rm area}^{\rm CNM}(L)\propto L^{-0.5}$. Increasing the beam size by a factor of $\sim40$ (from 15~pc to 0.6~kpc) reduces the covering fraction by $\sim40^{-0.5}\simeq0.16$, i.e., a factor of $\sim6$.

The WNM is distributed on scales larger than $L$ and is largely unaffected. The total observed \hi\ column density can then be written as
\begin{equation}
N_{\rm \ion{H}{i}}^{\rm tot}(L)
\simeq
N_{\rm \ion{H}{i}}^{\rm WNM}
+
C_{\rm area}^{\rm CNM}(L)\,N_{\rm \ion{H}{i}}^{\rm CNM}.
\end{equation}
When deriving $f_{\rm CNM}$, the WNM is implicitly assumed to trace the same pencil beam as the CNM. At larger $L$, this assumption breaks down: much of the WNM emission originates outside the absorption sight line, leading to an overestimate of the WNM contribution and hence to an underestimate of $f_{\rm CNM}$ by a factor $C_{\rm area}^{\rm CNM}$.
Although this is a simplified model, the predicted factor of $\sim6$ reduction in $C_{\rm area}^{\rm CNM}$ is broadly consistent with the factor of $\sim4$ difference between our measurements and those of \citet{pingel2024}.

A secondary effect arises in the derivation of the CNM spin temperature. If the emission spectrum includes WNM emission not present along the absorption sight line, the inferred CNM brightness temperature is biased low, leading to lower $T_s$. Since $N_{\rm \ion{H}{i}}^{\rm CNM}$ scales linearly with $T_s$, this effect further reduces the inferred CNM column density.
\erwin{The similarity between our results and those of \citet{pingel2024}, suggest both are affected by the same bias.}

Finally, the CNM linewidths $\Delta v_{\rm FWHM}$ are comparable to those found by \citet{pingel2024}, and the spin temperature ranges largely overlap. This indicates that both studies probe the same class of cold neutral structures. The differences in $f_{\rm CNM}$ and $N_{\rm \ion{H}{i}}^{\rm CNM}$ can therefore be attributed primarily to beam-averaging effects and the sampling of a structured CNM, rather than to intrinsic differences in the gas properties.
\erwin{In addition, the more modest velocity resolution (compared to most MW absorption studies) likely limits the ability to disentangle CNM structures blended in velocity.}

\subsection{Detection rates}

If we assume that the $2 \times 10^{20}$ \cm column density threshold determines whether absorption occurs, we can estimate the fraction of the area of each \hi\ disk above this limit. Using the \rzero zeroth-moment maps (with a sensitivity of $N_{\rm \ion{H}{i}} \sim 6 \times 10^{19}$ \cm) to define the disk extent, we find that 57~percent of the area in NGC~289 lies above this threshold, with an average disk radius of 54~kpc. For NGC~7424, the corresponding fraction is 54~percent with an average radius of 18~kpc. For the full MHONGOOSE sample, the average fraction is 56 percent. Thus, for sight lines intersecting the \hi\ disk, absorption would be expected in only slightly more than half of the cases.

We can compare our detection rate with that found by \citet{pingel2024} for NGC~6822. Their detection rate is also low, with 2 detections in 18 sight lines (11 percent). This is consistent with our rate of 3/56 (5 percent), or 3/31 (10 percent) for the clean sample. \citet{pingel2024} note that this is significantly lower than the 38 percent detection rate found by \citet{dempsey2022} for the SMC, despite both galaxies being gas-rich, low-mass, and low-metallicity systems. They further mention that all SMC sight lines lie within the $6 \times 10^{20}$ \cm contour. Applying this threshold to NGC~6822 increases the detection rate to 2/8 (25 percent).

Applying the same threshold to the MHONGOOSE data yields a detection rate of 3/20 (15 percent) for all continuum sources behind regions with $N_{\rm \ion{H}{i}} > 6 \times 10^{20}$ \cm. Excluding sources associated with H$\alpha$ increases this to 3/11 (27 percent, and restricting further to the clean sample gives 3/5 (60 percent). The inferred detection rate therefore depends strongly on the sample definition.
The $6 \times 10^{20}$ \cm contour encloses 0.05~deg$^2$ in NGC~6822 and 0.11~deg$^2$ across all MHONGOOSE galaxies. The two detections in NGC~6822 and three in MHONGOOSE imply comparable detection rates per unit area.

These results show that the detection rate does not depend on sensitivity alone. This complicates extrapolation to blind \hi\ absorption surveys at higher redshift.
A key difference is the flux density of the background sources. Our brightest continuum source has a flux of $\sim60$~mJy, whereas many absorption surveys use sources more than an order of magnitude brighter. This significantly improves optical-depth sensitivity.

As an example, \citet{gupta2010} study absorption in quasar--galaxy pairs out to $z \sim 0.1$. Restricting to the 16 systems with impact parameters $<20$~kpc (i.e.\ likely intersecting the main \HI\ disk; cf.\ \citealt{reeves2015,reeves2016}), $\sim80$ percent have continuum flux densities $>0.1$~Jy and $\sim35$ percent exceed 0.5~Jy. Of these, six ($\sim38$ percent) show \hi\ absorption.
For the sight line towards SUMSS~J225729$-$410241, we achieve our best optical-depth sensitivity of $\sigma_\tau = 0.014$. Assuming a FWHM of 10 \kms, as in \citet{gupta2010}, this corresponds to a $3\sigma$ upper limit of $\int \tau\, dv = 0.45$ \kms. Comparing with the values reported by \citet{gupta2010}, we would have detected only 1--2 of their absorption systems at this sensitivity. This corresponds to a detection rate of 13--26 percent, significantly lower than their observed rate, and is entirely due to the lower continuum flux densities in our sample.
Wide survey areas increase the number of background continuum sources available for absorption searches, but optical-depth sensitivity ultimately determines the fraction of detectable systems (modulo the distribution of absorbing \hi\ within galaxy disks).

\section{Summary}

We searched for H\,\textsc{i} absorption toward continuum sources behind the \HI\ disks of the MHONGOOSE galaxies and detect absorption along three lines of sight in two systems, NGC~289 and NGC~7424. With distances an order of magnitude larger than those of Local Group galaxies, the high angular resolution and sensitivity of the MHONGOOSE data allow the combined emission--absorption method -- traditionally applied in the MW and Local Group  -- to be extended to a new distance regime.

The detections occur along lines of sight with both high continuum peak flux density and high foreground \HI\ column density, while other sight lines show no absorption. This can be understood as the combined effect of two factors. First, below a foreground column density of $\sim 2\times10^{20}$ \cm, the formation of a substantial CNM component is not expected \citep{kanekar2011}. Second, optical-depth sensitivity imposes a strict detection limit: given the sensitivity of our data and the continuum flux densities, many sight lines are either too faint or have insufficient foreground \HI\ to produce detectable absorption.

We compare the derived CNM properties with those from the \HI\ absorption study of NGC~6822 by \citet{pingel2024} which has comparable sensitivity and resolution. We find CNM column densities and fractions that are lower by a factor of $\sim4$. We argue that this difference is not intrinsic, but arises from the larger physical scales over which the MHONGOOSE emission spectra are averaged. As a result, the emission spectrum is increasingly dominated by WNM gas unrelated to the narrow absorption sight line, leading to an overestimate of the WNM contribution and an underestimate of $f_{\rm CNM}$. The similarity in linewidths and spin temperatures between the samples indicates that both studies probe the same class of cold atomic structures, and that the observed differences are primarily due to resolution effects.

This study demonstrates that the emission--absorption method can be extended to much larger distances, provided that care is taken in constructing representative emission spectra. This has important implications for future facilities such as SKA-Mid and the Deep Synoptic Array (DSA; \citealt{hallinan2019}), which will offer beam sizes smaller by a factor of ${\sim}2$ for \HI\ emission compared to this study. This will improve the match between emission and absorption sight lines and reduce systematic biases in derived CNM fractions and spin temperatures. Their increased collecting areas will also enhance optical-depth sensitivity, reducing the reliance on bright background continuum sources. Combined with improved modeling of resolution effects on cold gas distributions, this will enable detailed studies of the CNM and WNM in more distant galaxies.

\begin{acknowledgements}
We thank the anonymous referee for the constructive comments. 
The MeerKAT
telescope is operated by the South African Radio Astronomy Observatory, which
is a facility of the National Research Foundation, an agency of the Department
of Science and Innovation. 
This work has received funding from the European Research Council (ERC) under the European Union’s Horizon 2020 research and
innovation programme (grant agreement No 882793 ‘MeerGas’). 
FMM carried out part of the research activities described in this paper with contribution of the Next Generation EU funds within the National Recovery and Resilience Plan (PNRR), Mission 4 - Education and Research, Component 2 - From Research to Business (M4C2), Investment Line 3.1 - Strengthening and creation of Research Infrastructures, Project IR0000034 –``STILES - Strengthening the Italian Leadership in ELT and SKA''.
\end{acknowledgements}

\bibliographystyle{aa}
\bibliography{absorption}

\begin{appendix}
\onecolumn
\section{The brightest non-detections}
Here we present the spectra of the non-detections towards the 14 brightest continuum sources listed in Table~\ref{tab:contcat}. The detection spectra are given in Fig.~\ref{fig:detectionspectra}. 
\begin{figure*}[h]
\centering
\includegraphics[width=0.6\textwidth]{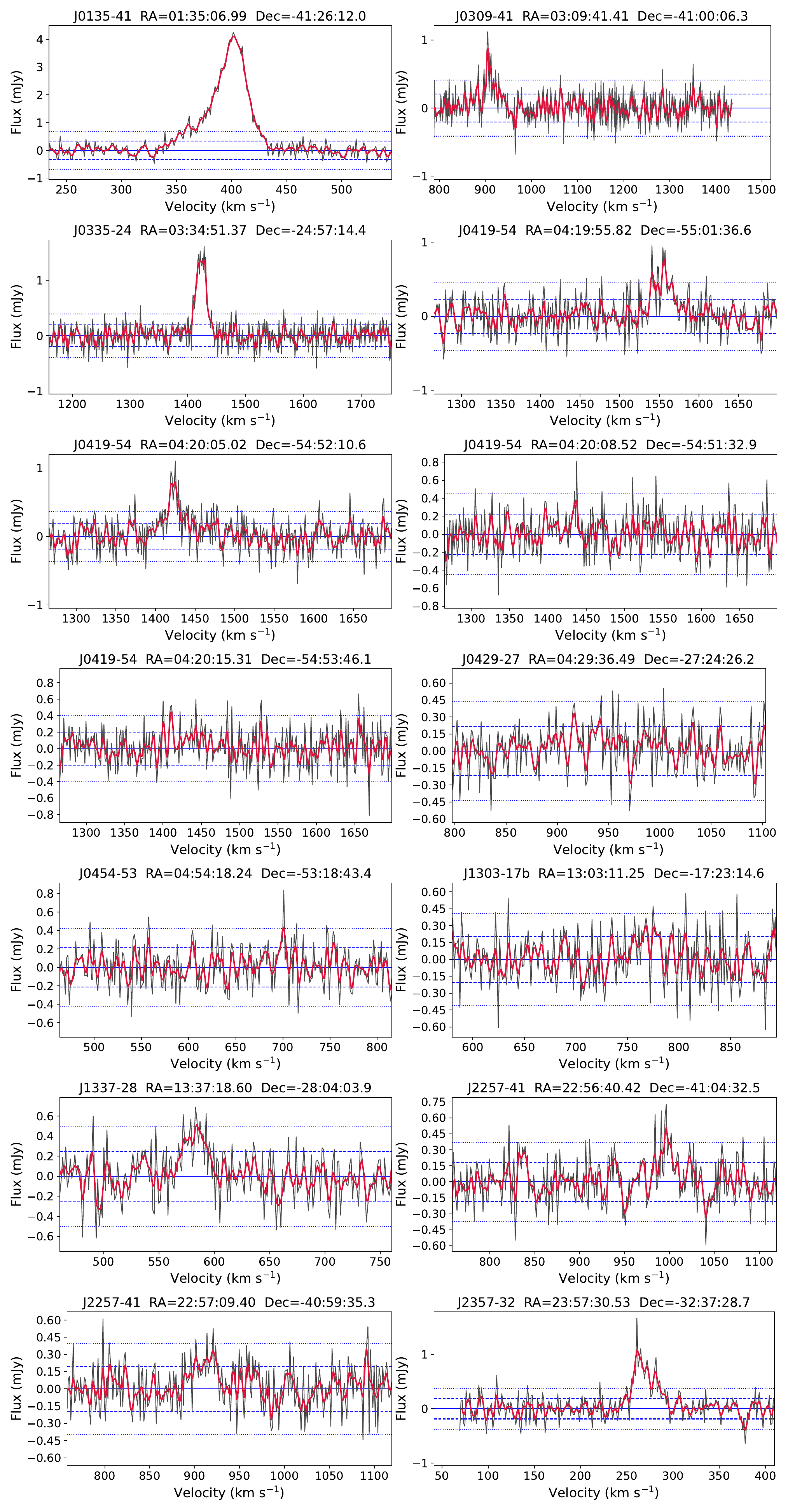}
\caption{Spectra of the 14 brightest non-detections as listed in Tab.~\ref{tab:contcat}. Lines and colours as in Fig.~\ref{fig:detectionspectra}.}
\label{fig:nodetections}
\end{figure*}



\end{appendix}
\end{document}